\documentclass[12pt]{article}
\usepackage{epsf}
\usepackage{amsmath,amssymb}
\usepackage{pdfpages}
\usepackage{graphicx}
\usepackage{comment}
\usepackage{tikz}
\usepackage{tikz-feynhand}
\usepackage[subrefformat=parens]{subcaption}
\usetikzlibrary{positioning}
\usepackage{physics}
\usepackage{slashed}
\usepackage{color}
\usepackage[driverfallback=dvipdfm]{hyperref}

\allowdisplaybreaks[1]

\begin{document}

\newcommand{\lsim}{\stackrel{<}{_\sim}}
\newcommand{\gsim}{\stackrel{>}{_\sim}}

\newcommand{\rem}[1]{\textcolor{magenta}{$\spadesuit$\bf #1$\spadesuit$}}

\renewcommand{\theequation}{\thesection.\arabic{equation}}

\renewcommand{\thefootnote}{\fnsymbol{footnote}}
\setcounter{footnote}{0}

\begin{titlepage}

\def\thefootnote{\fnsymbol{footnote}}

\begin{center}

\hfill May, 2022\\

\vskip .75in

{\Large \bf

  Leptophilic Gauge Bosons
  \\[2mm]
  at Lepton Beam Dump Experiments
  \\

}

\vskip .5in

{\large Takeo Moroi and Atsuya Niki}

\vskip 0.25in

{\em Department of Physics, University of Tokyo,
Tokyo 113-0033, Japan}

\end{center}
\vskip .5in

\begin{abstract}

  It has been recently known that we can use beams of future lepton
  colliders, the International Linear Collider (ILC), the Compact Linear
  Collider (CLIC), and the muon collider, for beam dump experiment if
  a shield and a detector are installed behind the beam dump.  We
  study the prospect of searching for leptophilic gauge bosons (LGBs)
  in association with $U(1)_{L_e-L_\mu}$, $U(1)_{L_e-L_\tau}$, and
  $U(1)_{L_\mu-L_\tau}$ gauge symmetries at such lepton beam dump
  experiments.  We perform a detailed calculation of the event rates of
  the LGB events, taking into account bremsstrahlung and
  pair-annihilation processes.  We show that the lepton beam dump
  experiments at future lepton colliders can reach parameter regions
  which are not been covered.

\end{abstract}

\end{titlepage}

\renewcommand{\thepage}{\arabic{page}}
\setcounter{page}{1}
\renewcommand{\thefootnote}{\#\arabic{footnote}}
\setcounter{footnote}{0}

\setlength{\feynhandlinesize}{1pt}

\section{Introduction}
\label{sec:intro}
\setcounter{equation}{0} 

While the standard model (SM) successfully explains most of the
electroweak-scale phenomena, it is widely believed that there exists
physics beyond the standard model (BSM).  To search for the BSM
particles, the high energy collider experiments with lepton beam, such
as the International Linear Collider (ILC) \cite{Behnke:2013xla,
  Baer:2013cma, Adolphsen:2013jya, Adolphsen:2013kya, Behnke:2013lya,ILCInternationalDevelopmentTeam:2022izu},
the Compact Linear Collider (CLIC) \cite{Roloff:2018dqu}, and the muon
collider \cite{Delahaye:2019omf}, are highly appreciated as prominent
candidates of future collider experiments.
One advantage of lepton colliders is that elementary processes at the
interaction point and the background are well understood, so the
lepton collider experiments are suited for the precision test of the
high energy physics like Higgs properties.

In recent years, it has been pointed out that the high-energy lepton
collider experiments can be also used for beam dump experiments, a
kind of fixed target experiment \cite{Kanemura:2015cxa,
  Cesarotti:2022ttv}. In this experiment, beam is injected into a beam
dump, then new particles are produced through the interactions between
the beam and the material in the beam dump. Since one of the initial
states is a fixed target, the luminosity is greatly increased and the
beam dump experiment has an advantage to search for feebly interacting
particles compared to the collider experiments. In addition, the beam
dump experiment at the future lepton collider has several advantages
compared to fixed target experiments in the past: high energy beam
($>\mathcal{O}(100)\ \rm{GeV}$) and a large amount of the
initial-state leptons.  Due to these advantages, it has been shown
that the beam dump experiment using the ILC beam can cover the
parameter regions of long-lived BSM particles where the previous
experiments could not explore \cite{Kanemura:2015cxa, Sakaki:2020mqb,
  Asai:2021ehn, Asai:2021xtg}.

Among various possibilities of BSM physics, in this paper, we
concentrate on gauge bosons in association with new $U(1)$ gauge
symmetries coupled to the difference of the lepton-family numbers,
$U(1)_{L_e-L_\mu}$, $U(1)_{L_e-L_\tau}$, and $U(1)_{L_\mu-L_\tau}$
\cite{Foot:1990mn, He:1990pn, He:1991qd, Foot:1994vd}.  These new
$U(1)$ gauge symmetries can be introduced without quantum anomaly.  We
call the gauge bosons of our interest as leptophilic gauge bosons
(LGBs). The phenomenology of leptophilic $U(1)$ gauge symmetry have
been intensively discussed; relevant subjects include neutrino physics
\cite{Araki:2012ip, Heeck:2014sna, Asai:2017ryy, Asai:2018ocx,
  Asai:2019ciz}, muon anomalous magnetic moment
\cite{Ma:2001tb,Baek:2001kca}, dark matter
\cite{Foldenauer:2018zrz,Holst:2021lzm,Drees:2021rsg}, and so
on. These LGBs can be long-lived in some parameter regions so they may
be detected by the beam dump experiment.  Ref.\ \cite{Asai:2021xtg}
considered the search for LGBs in the ILC beam dump experiment and
show that the ILC beam dump experiment can explore the parameter
region which is still viable.  In particular, for the case of
$U(1)_{L_\mu-L_\tau}$, the ILC beam dump experiment can cover part of
the regions where the LGB alleviates the Hubble tension
\cite{Escudero:2019gzq}.  In Ref.\ \cite{Asai:2021xtg}, only the
effect of the $e^-$ beam particle was taken into account. However,
there can occur scattering processes using the secondary $e^{\pm}$'s
and $\mu^{\pm}$'s.  In addition, the LGB can be produced via the
annihilation process between the energetic beam or secondary $e^+$ and
the atomic $e^-$.  These processes, which also contribute to the
production of the LGB, are considered in the present study.

In this paper, we consider the possibility to search for the LGBs with
the beam dump experiments using energetic beams of lepton colliders.
We study in detail the production processes of LGBs as well as the
detection efficiency and obtain the discovery reach.  We take into
account the production processes of LGBs which have not been
considered in the previous analysis: the secondary $e^\pm$ and
$\mu^\pm$ bremsstrahlung process and the positron $e^+$ annihilation
process.  The cross section of the annihilation process can be larger
than that of the bremsstrahlung process in some parameter space, so the sensitivity can be
improved by this process.  Furthermore, since the LGB of
$U(1)_{L_\mu-L_\tau}$ model does not couple to $e^\pm$ at tree
level, the the effects of secondary $\mu^\pm$ may be non-negligible.
We also consider the search for LGBs using the muon beam of the muon
colliders.

The organization of this paper is as follows.  We introduce the
$U(1)_{L_i-L_j}$ ($i,j=e,\mu,\tau$) models in Sec.\ \ref{sec:model}
and the beam dump experiment in Sec.\ \ref{sec:bd}.  Then we explain
the calculation of the  event rate in Sec.\ \ref{sec:event rate}. 
We
show the discovery reaches for the each model in the lepton beam dump
experiments in Sec.\ \ref{sec:discovery} and Sec.\ \ref{sec:muon_bd}.
Sec.\ \ref{sec:summary} is devoted for summary and discussion.

\section{Model}
\label{sec:model}
\setcounter{equation}{0}

We consider the $U(1)_{L_i-L_j}\ (i,j=e,\mu,\tau)$ models, where $L_i$
is the lepton $i$ number. The charge assignment for the lepton $\ell$
is
\begin{align}
  Q_\ell
  \equiv
  \left\{ \begin{array}{ll}
    1 & :\ell=i , \\[1mm]
    -1 & :\ell=j,\\
    0 & :\textrm{otherwise}.
  \end{array} \right.
\end{align}
These $U(1)$ extensions are gauge anomaly free without new particles,
and we do not introduce new matter particles charged under the
$U(1)_{L_i-L_j}$ symmetry for simplicity.\footnote
{When we introduce new particle which is charged under the
  $U(1)_{L_i-L_j}$ symmetry and singlet of the SM symmetries, this new
  particles can be a dark matter candidate
  \cite{Foldenauer:2018zrz,Holst:2021lzm,Drees:2021rsg}. The existence
  of this new particle may affect the sensitivity of the LGB in the
  beam dump experiment because the LGB may decay into this new
  particle, which cannot be detected. To search for this kind of
  particles in the dark sector, we should use the other type of the
  fixed target experiment setup, like LDMX \cite{LDMX:2018cma}.}
Furthermore, we assume that the $U(1)_{L_i-L_j}$ symmetry is spontaneously
broken and that the LGB is massive; the Higgs field responsible for the
spontaneous breaking of $U(1)_{L_i-L_j}$ is assumed to be heavy enough
so that it does not affect the following discussion. Then the relevant
part of the Lagrangian for our study is given by
\begin{align}
  \mathcal{L} =& -\frac{1}{4}F_{\mu\nu}F^{\mu\nu} -\frac{1}{4}X_{\mu\nu}X^{\mu\nu}-\frac{\epsilon_0}{2}X_{\mu\nu}F^{\mu\nu}+\frac{1}{2}m_{A'}^2A'_{\mu}A'^{\mu} + eA_{\mu}J^{\mu}_{EM} \nonumber\\
  &+ g'A'_{\mu}J^{\mu}_{DS}
  + g'A'_{\mu}\sum_{\ell=e,\mu,\tau}Q_\ell\qty(\bar{\ell}\gamma^{\mu}\ell+\bar{\nu_\ell}\gamma^{\mu}P_{L}\nu_{\ell}) + {\rm \cdots},
  \label{lag_lp}
\end{align}
where $P_L\equiv\frac{1}{2}(1-\gamma_5)$,  $A_\mu$ ($A'_\mu$) is the photon (the LGB) and $F_{\mu\nu}$
($X_{\mu\nu}$) is its field strength. 

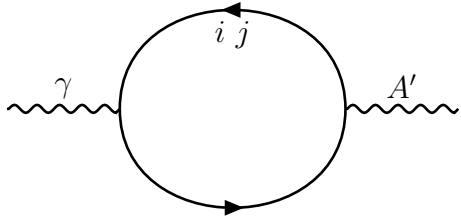
\begin{figure}[t]
  \centering
  \begin{tikzpicture}
    \begin{feynhand}
      \vertex (i) at (-3,0);
      \vertex (f) at (3,0);
      \vertex (m1) at (-1.5,0);
      \vertex (m2) at (1.5,0);
      \propag [pho] (i) to [edge label = $\gamma$](m1);
      \propag [bos] (m2) to [edge label =$A'$](f);
      \propag [fer] (m1) to [half right,looseness=1.5](m2);
      \propag [fer] (m2) to [half right, edge label = $i$\ $j$,looseness=1.5](m1);
    \end{feynhand}
  \end{tikzpicture}
  \caption{1-loop induced kinetic mixing between photon and
    leptophilic gauge boson via exchanging the corresponding charged
    leptons.}
  \label{fig:kinetic_loop}
\end{figure}

In Eq.\ \eqref{lag_lp}, the  third term is the tree-level
kinetic mixing term.  There is also a 1-loop contribution to the kinetic mixing, as is shown
in Fig. \ref{fig:kinetic_loop}.  Then, the effective mixing parameter,
which is used to calculate the decay rate of the LGB, is given by
\begin{align}
  \epsilon \equiv \epsilon_0 + \Delta \epsilon,
\end{align}
where 
\begin{align}
  \Delta\epsilon =
  \frac{eg'}{2\pi^2}\int^{1}_{0}\dd x\ x(1-x)
  \ln\frac{m_{j}^2-m_{A'}^2x(1-x)}{m_{i}^2-m_{A'}^2x(1-x)}.
\end{align}
After diagonalizing the kinetic terms of gauge bosons, the Lagrangian
contains the following interaction:
\begin{align}
  \mathcal{L} \ni 
  - \epsilon e A'_{\mu}J^{\mu}_{EM}.
\end{align}
In order to reduce the unknown parameter, we set the tree-level
kinetic mixing parameter to be zero, $i.e.$, $\epsilon_0=0$.  Then the
interactions between the LGB and the electromagnetic current suffer
the 1-loop suppression.

The interactions between LGB and SM particles can be denoted as
\begin{align}
  \mathcal{L} \ni
  - \sum_\ell \left (
  g_\ell\bar{\ell}\slashed{A'}\ell
  + g_{\nu_\ell} \bar{\nu}_\ell \slashed{A'} P_L \nu_\ell
  \right),
\end{align}
where
\begin{align}
  -g_i = g_j = g', ~~~ g_{\ell\neq i,j} = \epsilon e, ~~~
  -g_{\nu_i} = g_{\nu_j} = g',~~~ g_{\nu_{\ell\neq i,j}} = 0.
\end{align}
In the case of the $U(1)_{L_e-L_\mu}$ and $U(1)_{L_e-L_\tau}$ models, the LGB couples with $e$ via the gauge interaction. On the other hand, in the case of the $U(1)_{L_\mu-L_\tau}$ model, 
the LGB couples with $\mu$ and $\tau$ flavor leptons at tree level and with $e$ at only loop level.

\begin{figure}
  \centering
  \begin{tabular}{cc}
    \begin{minipage}{0.5\hsize}
      \centering
      \includegraphics[width=8.5cm, angle=0]{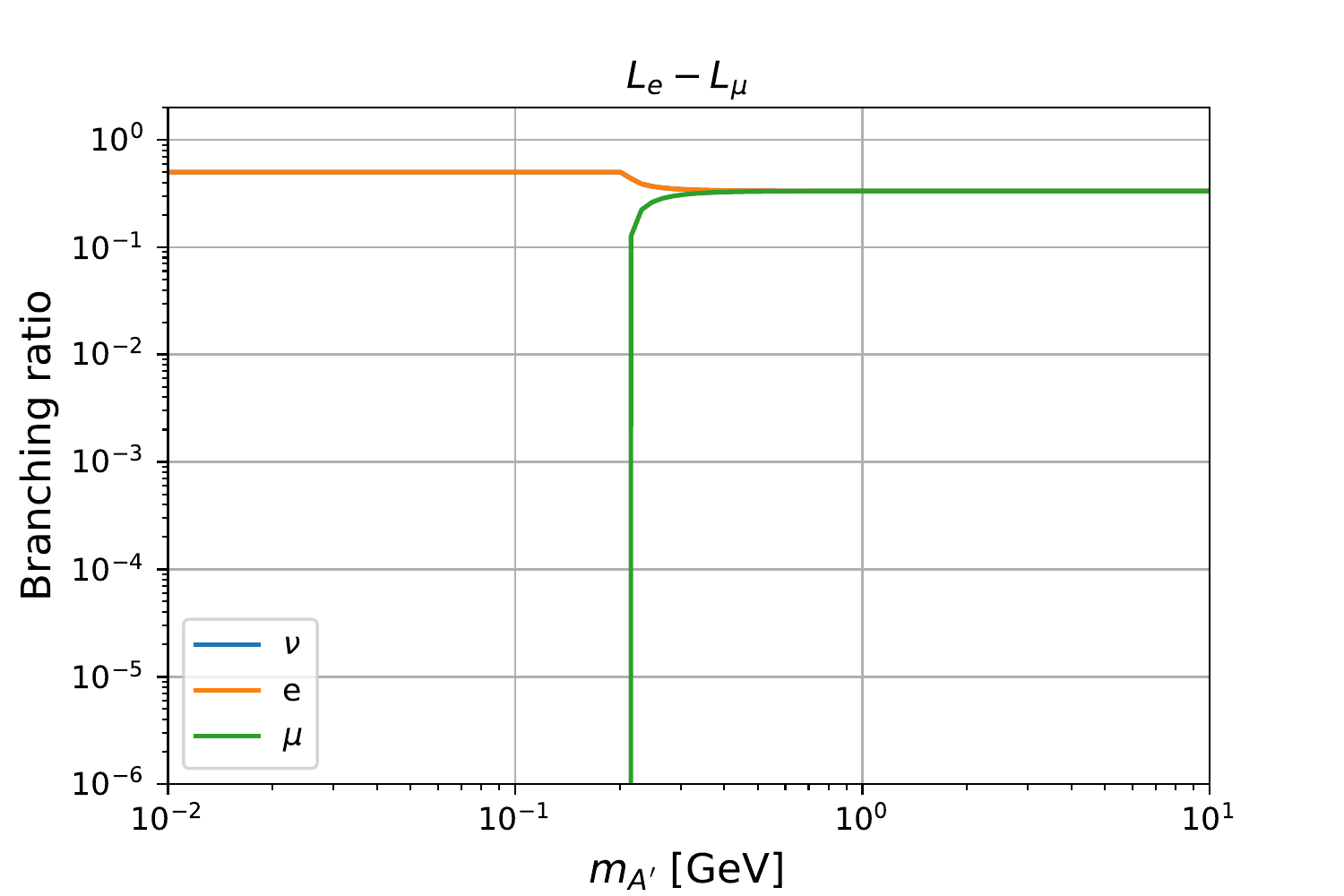}
      \hspace{2cm}
  \end{minipage}
    \begin{minipage}{0.5\hsize}
      \centering
      \includegraphics[width=8.5cm, angle=0]{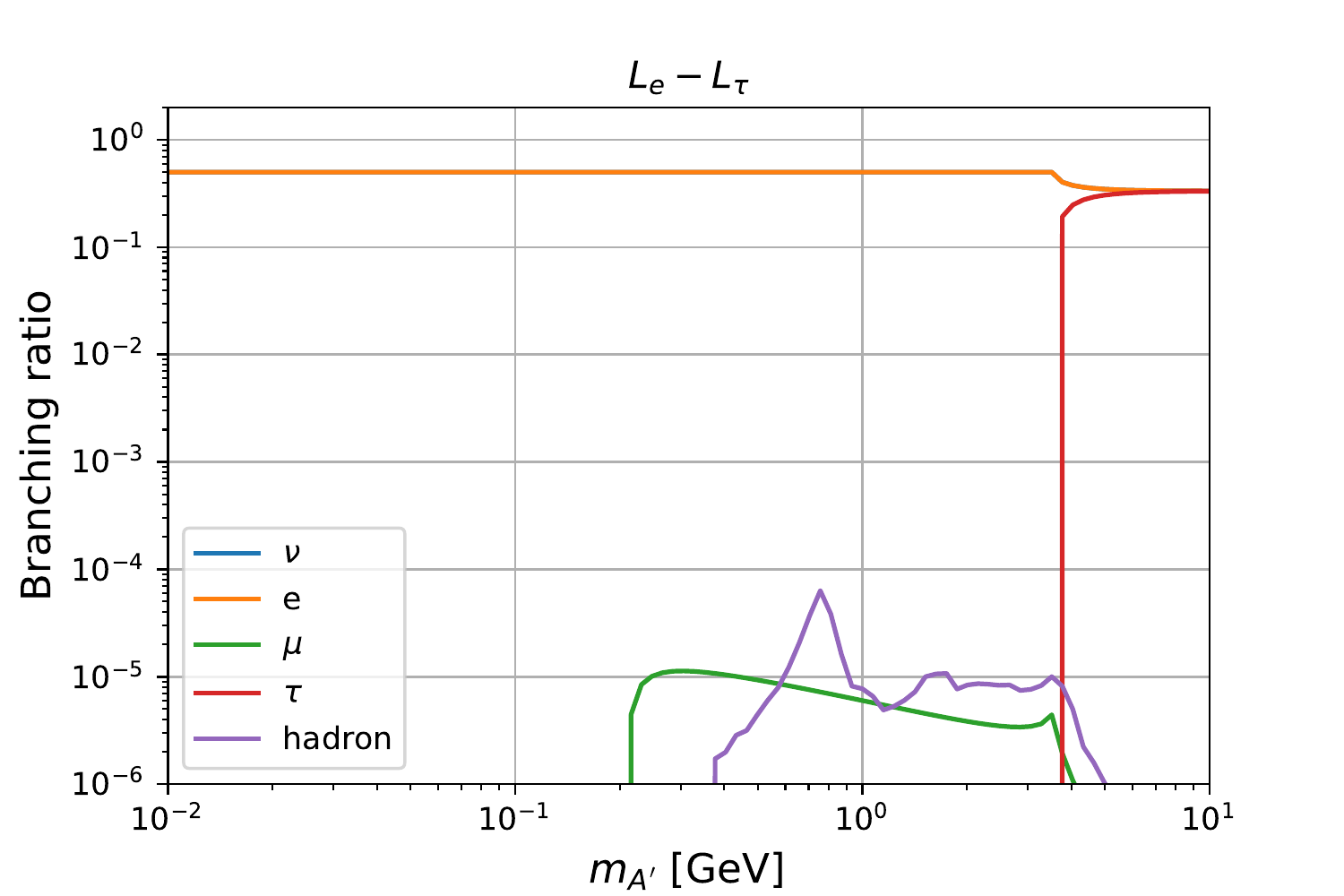}
      \hspace{2cm}
  \end{minipage}
    \end{tabular}
  \caption{The branching ratios for the $U(1)_{L_e-L_\mu}$ gauge boson (left) and the $U(1)_{L_e-L_\tau}$ gauge boson (right).  (Notice that the branching ratios for the $e^\pm$ and the neutrino final states are very close and hence they are indistinguishable in the fitures.)}
  \label{fig:branch_emu}
  \vspace{7mm}
  \begin{tabular}{cc}
  
  \begin{minipage}{0.5\hsize}
      \centering
      \includegraphics[width=8.5cm, angle=0]{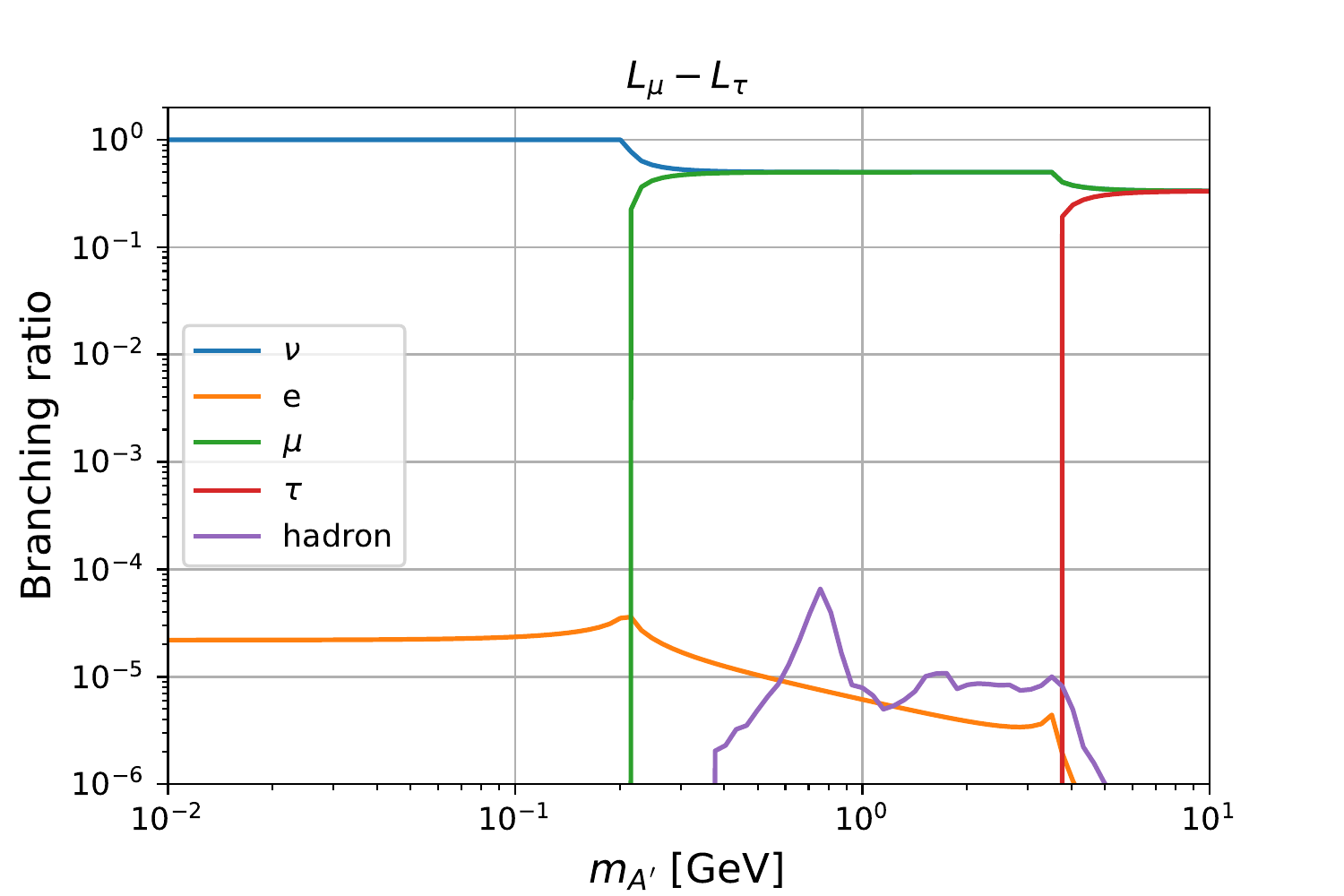}
      \hspace{2cm}
      \subcaption{Branching Ratio}
  \end{minipage}
  
  \begin{minipage}{0.5\hsize}
      \centering
      \includegraphics[width=8.5cm, angle=0]{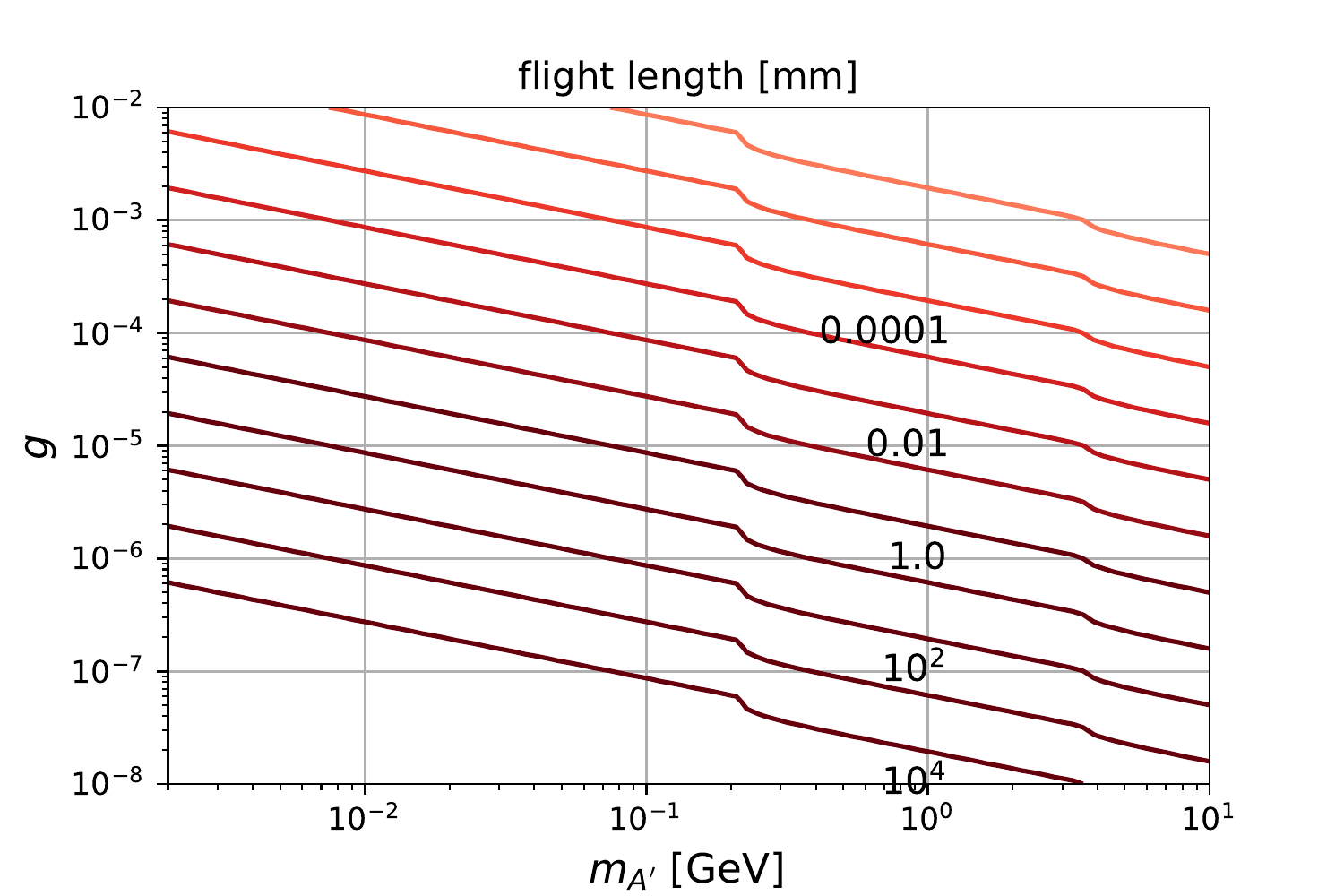}
      \hspace{2cm}
      \subcaption{Decay Length $c\tau$ [mm]}
  \end{minipage}
  
  \end{tabular}
  \caption{The decay properties of the $U(1)_{L_\mu-L_\tau}$ gauge boson. (a) the branching ratio (b) the decay length $c\tau$, where $\tau$ is the lifetime.}
  \label{fig:branch_mut}
\end{figure}

Through these interactions, the LGB can decay into the SM particles when the decay mode is kinematically allowed. The LGB's partial decay rates to the charged lepton pair are given by
\begin{align}
  \Gamma(A'\rightarrow \ell^+\ell^-) 
  =
  \left\{ \begin{array}{ll}
      \frac{g'^2}{12\pi}m_{A'}\sqrt{1-\frac{4m_\ell^2}{m_{A'}^2}}\qty(1+\frac{2m_l^2}{m_{A'}^2}) & :\ell=i,j,\\[1mm]
      \frac{(\epsilon e)^2}{12\pi}m_{A'}\sqrt{1-\frac{4m_l^2}{m_{A'}^2}}\qty(1+\frac{2m_l^2}{m_{A'}^2}) &: \textrm{otherwise},
  \end{array}\right.
\end{align}
while the partial decay rates to neutrino pair is given by
\begin{equation}
  \Gamma(A'\rightarrow \nu_\ell\bar{\nu_\ell}) 
  =
  \left\{ \begin{array}{ll}
      \frac{g'^2}{24\pi}m_{A'} & :\ell=i,j,\\[1mm]
      0 &: \textrm{otherwise}.
  \end{array}\right.
\end{equation}
Furthremore, the partial decay rate to hadron can be expressed as
\begin{equation}
  \Gamma (A'\rightarrow {\rm hadron}) = \Gamma(A'\rightarrow \mu^+\mu^-)R(m_{A'}),
\end{equation}
where $R$ is the $r$-ratio \cite{Ezhela:2003pp,Zyla:2020zbs}. The
branching ratio of the $U(1)_{L_i-L_j}$ gauge bosons are shown in
Fig.\ \ref{fig:branch_emu} and Fig.\ \ref{fig:branch_mut}. In the case of the $U(1)_{L_e-L_\mu}$ and $U(1)_{L_e-L_\tau}$ models, the LGB decays into $e^+e^-$ pair dominantly as well as neutrino pair and the branching ratio into $e^+e^-$ pair is $\mathcal{O}(0.1)$. 
On the other hand, the LGB of the $U(1)_{L_\mu-L_\tau}$ model couples
with muon and tau flavor at tree level but that it interacts with
electron flavor and quarks only via the loop effect. So the branching ratio
of the decay modes to electron pair and hadrons are suppressed. In
particular, when the LGB mass is less than twice of the muon mass, the
possible final states of the LGB decay are $e^+e^-$ pair and neutrino
pair. As is explained, the decay rate into $e^+e^-$ pair is
suppressed and most of the decay products are invisible neutrinos in
this mass range. This makes it difficult to find light LGB in
$U(1)_{L_\mu-L_\tau}$ model.

\section{Beam Dump Experiment}
\label{sec:bd}
\setcounter{equation}{0}

In this section, we explain the experimental setup of the lepton beam
dump experiment. In $e^+e^-$ linear colliders such as the ILC and the CLIC, the
beams passing through the interaction point are expected to be dumped
into the beam dumps. Then, the beam dump can be used as a target for
the beam dump experiment if additional equipment is installed behind
the beam dump. In the following, we consider such a possibility.

\begin{figure}[t]
  \centering
  \includegraphics[width=17cm,angle=0]{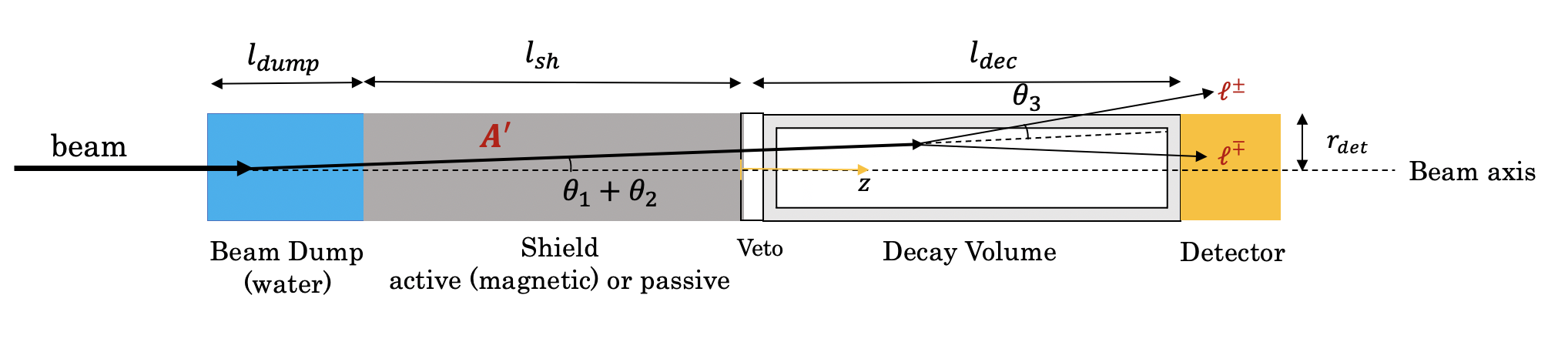}
  \caption{Rough sketch of the experimental setup of the lepton beam dump experiment. $A'$ is the LGB and $l^\pm$ is its decay particles. $r_{det}$ is the radius of the detector. $A'$ is produced at the beam dump ($A'$ can be also produced in the shield by the secondary muon) and then it travels to the decay volume. In the decay volume, $A'$ decays into a lepton pair $\ell^+\ell^-$.
  We observe this lepton pair at the detector.}
  \label{fig:BD_setup}
\end{figure}

The sketch of the setup is shown in Fig. \ref{fig:BD_setup}.
Behind the beam dump, a shield, veto and the decay volume with
particle detector are installed; the lengths of the beam dump, the
shield, and the decay volume are denoted as $L_{dump}$, $L_{shield}$, and
$L_{dec}$, respectively.  The shield is intended to eliminate muon and
other backgrounds.\footnote
{We can consider two types of shields, active shield or passive
  shield. The active shield is the SHiP-like shield
  \cite{SHiP:2017wac}, in which we use the magnetic field to remove
  the muons. The passive shield is just an enough-long lead shield to
  stop the muons. Using this shield the secondary muons can also
  produce the new gauge bosons.}
Once a LGB is produced in the beam dump, it may pass through the beam
dump and the shield, reaching to the decay volume. If the LGB
decays into visible particles in the decay volume, they may be
observed by the detector.  Hereafter, we assume that charged SM
particles are fully blocked by the shield and that the background is
negligible.

Before closing this section, we comment on the mean free path of the
LGB in the beam dump and the shield.  In our analysis, we consider the
case that the beam energy is a much larger than the LGB mass $m_{A'}$.
In this case, the mean free path of the LGB can be obtained by
rescaling that of the photon.  When the LGB energy is larger than
$1\ \rm{GeV}$, the LGB mean free path is estimated to be
$L\sim\frac{e^2}{g'^2}\ \rm{cm}$ in the lead \cite{Zyla:2020zbs}. As
we will see, we will investigate the region of $g'<10^{-4}$, for which
the mean free path is much larger than $100\ \rm{m}$.  Then, with the
experimental setup of our choice, we can safely neglect the
interaction of the LGB in the beam dump and shield.

\section{Event Rate}
\label{sec:event rate}
\setcounter{equation}{0}

In this section, we explain how to calculate the event rate in the
beam dump experiment.  The event rate $N$ is expressed as
\begin{equation}
  N = N_l N_{target} \sigma \mathcal{A},
  \label{eq:rough_N}
\end{equation}
where $N_l$ is the number of the leptons injected into the beam dump per time, $N_{target}$
is the effective number of partcles in the target, $\sigma$ is the
cross section for the production process and $\mathcal{A}$ is the acceptance of
the detection. More details will be discussed in the following.

\subsection{Production process}

\begin{figure}[t]
  \centering
  \begin{tabular}{cc}
  \begin{minipage}{5cm}
  \begin{tikzpicture}
      \begin{feynhand}
          \vertex [particle] (i1) at (-2,2) {$e, \mu$};
          \vertex [particle] (i2) at (-2,0) {$Z$};
          \vertex [particle] (f1) at (2,3) {$A^{'}$};
          \vertex [particle] (f2) at (2,2) {$e,\mu$};
          \vertex [particle] (f3) at (2,0) {$Z^{'}$};
          \vertex (w1) at (0,2.2);
          \vertex (w2) at (0,2);
          \vertex (w3) at (0,0) ;
          \propag [fer] (i1) to (w2);
          \propag [fer] (i2) to (w3);
          \propag [fer] (w2) to (f2);
          \propag [bos] (w1) to (f1);
          \propag [fer] (w3) to (f3);
          \propag [bos] (w2) to [edge label=$\gamma$](w3);
      \end{feynhand}
  \end{tikzpicture}
  \subcaption{bremsstrahlung}
  \end{minipage}

  \hspace{2cm}

  \begin{minipage}{5cm}
  \begin{tikzpicture}
      \begin{feynhand}
          \vertex [particle] (i1) at (-2,3) {$e^+$};
          \vertex [particle] (i2) at (-2,0) {$e^-$};
          \vertex [particle] (f1) at (2,1.5) {$A^{'}$};
          \vertex (w1) at (0,1.5);
          \propag [fer] (w1) to (i1);
          \propag [fer] (i2) to (w1);
          \propag [bos] (w1) to (f1);
      \end{feynhand}
  \end{tikzpicture}
  \subcaption{annihilation}
  \end{minipage}

  \end{tabular}
  \caption{Production processes in the lepton beam dump experiment. $A'$ is a LGB and $Z,Z'$ are the nucleus.}
  \label{fig:production_process}
\end{figure}
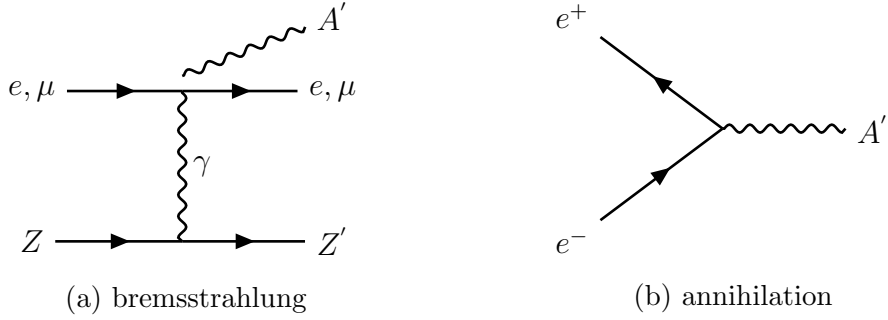

In this subsection, we discuss the production processes and the cross
sections of the LGBs. In the $e^\pm$ beam dump experiment, the
candidates of the initial-state particle are the beam $e^\pm$ and the
secondary $e^\pm$, $\mu^\pm$, and $\gamma$. For the LGB production,
the bremsstrahlung process by the charged particle and the
annihilation process by the incoming positron and the atomic electron
are the dominant production processes (see
Fig.\ \ref{fig:production_process}).

\subsubsection{Bremsstrahlung}
The bremsstrahlung is the radiation process from the charged particles
propagating in the matter. The diagram is shown in
Fig. \ref{fig:production_process} (a).  We calculate the cross section
using the improved Weizsacker-Williams approximation
\cite{Kim:1973he,Tsai:1973py,Tsai:1986tx}. The cross section with the
incoming charged particles $\ell^\pm (=e,\mu)$ is\footnote{The
  approximation for the cross section is a little bit different from
  the $e^\pm$ bremsstrahlung cross section
  \cite{Kirpichnikov:2021jev}. This is because $m_\mu$ is not
  necessary smaller than the LGB mass $m_{A'}$.}
\begin{align}
  \frac{\dd\sigma(\ell+Z\rightarrow \ell+Z'+A')}{\dd x\dd \cos\theta_{A'}} = &\frac{2\alpha^2g_\ell^2}{\pi}\chi\beta_{A'}E_\ell^2x
  \left[\frac{1-x+\frac{1}{2}x^2}{U^2}\right.\nonumber\\
  &\left.+\frac{(1-x)^2(m_{A'}^2+2m_l^2)}{U^4}\qty(m_{A'}^2-\frac{x}{1-x}U+\frac{x^2}{1-x}m_\ell^2)\right],
  \label{eq:WW_cross}
\end{align}
where $E_{A'}\ (E_\ell)$ is the energy of the LGB (the charged particle
$\ell^\pm$), $\theta_{A'}$ is the angle of the LGB from the propagation
direction of $\ell^\pm$, and 
\begin{align}
  x &= \frac{E_{A'}}{E_\ell}\\
  \beta_{A'} &= \sqrt{1-\frac{m_{A'}^2}{E_{A'}^2}}\\
  U(x,\theta_{A'}) &\equiv E_\ell^2\theta_{A'}^2x+m_{A'}^2\frac{1-x}{x}+m_\ell^2x.
\end{align}
In addition, $\chi$ is the effective photon flux:
\begin{equation}
  \chi = \int^{\tilde{t}_{max}}_{\tilde{t}_{min}}\dd \tilde{t}\ \frac{\tilde{t}-\tilde{t}_{min}}{\tilde{t}^2}G_2(\tilde{t}),
\end{equation}
where the elastic and inelastic form factor 
of the target of the mass number $A$ is given by
\begin{equation}
  G_2 (\tilde{t})
  = \qty(\frac{a^2\tilde{t}}{1+a^2\tilde{t}})^2\qty(\frac{1}{1+\tilde{t}/d})^2Z^2+\qty(\frac{a'^2\tilde{t}}{1+a'^2\tilde{t}})^2\qty(\frac{1+(\mu_p^2-1)\tilde{t}/4m_p^2}{(1+\tilde{t}d')^4})^2Z,
\end{equation}
with $m_p$ being the proton mass, $a = 111Z^{-\frac{1}{3}}/m_e$,
$d=0.146\ {\rm{GeV}^2} A^{-\frac{2}{3}}$, $a' = 773Z^{-\frac{2}{3}}/m_e$, $d=0.71\ \rm{GeV}^2$, and
$\mu_p = 2.79$. We set $t_{min}\simeq \qty(\frac{m_{A'}^2}{2E_\ell})^2$
and $t_{max}\simeq m_{A'}^2+m_\ell^2$ \cite{Liu:2016mqv,Liu:2017htz}.

The cross section is enhanced in the collinear region $\theta_{A'}\sim
0$. Furthermore, the LGB is detected only when the LGB flies collinearly to
the beam axis. Due to these reasons, we 
concentrate on the case that 
$\theta_{A'}$ is very small.  Then, 
integrating the diffrential cross section for 
$\theta_{min}\leq\theta_{A'}\leq\theta_{max}$, we obtain
\begin{align}
  \frac{\dd\sigma}{\dd x}
  =&\frac{2\alpha^2 g_\ell^2}{\pi}\chi\beta_{A'}\left[\frac{1-x+\frac{1}{2}x^2}{E_\ell^2x}\frac{1}{2}\qty(\frac{1}{\theta_{min}^2+\eta}-\frac{1}{\theta_{max}^2+\eta}) \right. \nonumber\\
  & \left. +\frac{(1-x)^2(m_{A'}^2+2m_\ell^2)m_{A'}^2}{(E_\ell^2x)^3}\frac{1}{6}\qty(\frac{1}{(\theta_{min}^2+\eta)^3}-\frac{1}{(\theta_{max}^2+\eta)^3})\right.\nonumber\\
  &\left.-\frac{(1-x)x(m_{A'}^2+2m_\ell^2)}{(E_\ell^2x)^2}\frac{1}{4}\qty(\frac{1}{(\theta_{min}^2+\eta)^2}-\frac{1}{(\theta_{max}^2+\eta)^2})\right.\nonumber\\
  & \left. +\frac{(1-x)x^2(m_{A'}^2+2m_\ell^2)m_\ell^2}{(E_\ell^2x)^3}\frac{1}{6}\qty(\frac{1}{(\theta_{min}^2+\eta)^3}-\frac{1}{(\theta_{max}^2+\eta)^3})
  \right],
  \label{eq:brems_cross_overangle}
\end{align}
where $\eta=\frac{m_{A'}^2}{E_\ell^2}\frac{1-x}{x}+\frac{m_\ell^2}{E_\ell^2}$.

\subsubsection{Annihilation}

The incoming positron (beam particle or secondary particle) and the
electron in the atom can annihilate into the LGB, as is shown in
Fig. \ref{fig:production_process} (b). The cross section for the
annihilation process $e^+e^-\rightarrow A'\rightarrow \ell^+\ell^-$ is
\begin{align}
  \sigma^\ell_{ann} &= \frac{g_e^2g_\ell^2E_{CM}^2}{12\pi\qty[(E_{CM}^2-m_{A'}^2)^2+m_{A'}^2\Gamma_{A'}^2]}\sqrt{1-4\frac{m_\ell^2}{E_{CM}^2}}\qty(1+2\frac{m_\ell^2}{E_{CM}^2}),
\end{align}
where $E_{CM}$ is the center of mass energy.  If the LGB is
long-lived, the narrow-width approximation can be used:
\begin{align}
  \sigma^\ell_{ann}
  &= \frac{g_e^2g_\ell^2}{24}\frac{E_{CM}^2}{m_{A'}^2\Gamma_{A'}}\sqrt{1-4\frac{m_\ell^2}{E_{CM}^2}}\qty(1+2\frac{m_\ell^2}{E_{CM}^2})\delta(E_{CM}-m_{A'}).
\end{align}

\subsection{Experimental acceptance}

The LGB can be produced in the beam dump through the bremsstrahlung
and the annihilation process, as we discussed in the previous
subsection. However, not all of the LGBs can be detected. There are
two reasons: one is that not all of the LGB can travel through the
dump and the shield, and another is that the detector is assumed to be
installed in the very forward direction.\footnote
{We also check the
  situation when the detector is installed on every side of the decay
  volume and the sensitivity becomes better, as expected
  \cite{niki:2022m}.} 
We estimate the experimental acceptance (corresponding to the
detection efficiency of the LGB produced in the dump) as
\begin{equation}
  \mathcal{A} = \frac{1}{L_{A'}}
  e^{-(L_{dump}+L_{sh})/L_{A'}}
  \int_0^{L_{dec}} \dd z\,
  e^{-z/L_{A'}}\,
  \Theta(r_{det}-r_{\perp}(z)),
  \label{Acceptance}
\end{equation} 
where $z$ is the distance from the shield to the decay point of the
LGB, $r_{det}$ is the radius of the detector (as shown in
Fig. \ref{fig:BD_setup}), and
\begin{align}
  L_{A'}=\frac{p_{A'}}{\Gamma_{A'}m_{A'}}.
\end{align}
In addition, $r_{\perp}(z)$ is the distance of the final state
particle from the beam axis at the position of the detector. We adopt
an approximation that $r_{\perp}(z)$ is estimated by using typical
scattering and decay angles of the production; consequently, in our
analysis, $r_{\perp}(z)$ is calculated as a function of $z$ (see
below).  Then, the theta function 
in Eq.\ \eqref{Acceptance}
takes account of the angular
acceptance. Estimation of $r_{\perp}$ depends on the initial state
particle:
\begin{itemize}
  \item Incoming $e^\pm$: In this case, the production
    process mostly occurs at the front edge of the beam dump.  Then,
    the $r_{\perp}$ is estimated as
\begin{equation}
  r_\perp \simeq (\theta_1+\theta_2)(L_{dump}+L_{sh}+z)+(\theta_1+\theta_2+\theta_3)(L_{dec}-z),
  \label{eq:r_perp}
\end{equation}
where $\theta_{1}$ is the $e^\pm$ angle from the beam axis, $\theta_2$
is $A'$ angle from the propagation direction of $e^\pm$, and
$\theta_3$ is the angle between the propagation directions of $A'$ and
final state particle.\footnote
{Since the angles $\theta_{1,2,3}$ are independent, we may use the
  root mean square to estimate $r_\perp$.  Eq.\ (\ref{eq:r_perp})
  assumes the worst case when the final state flies away from the beam
  axis, so our estimatin is conservative. Ref. \cite{Asai:2021ehn}
  used the root mean square to estimate $r_\perp$. We have checked
  that, even if we use the root mean square, results do not change
  significantly. In our analysis, we use Eq.\ (\ref{eq:r_perp}) to
  simplify the calculation.}
Here the angles $\theta_{1,2,3}$ are approximated by the typical
values \cite{Bjorken:2009mm, Sakaki:2020mqb}:
\begin{align}
  \theta_1 &= 16\ {\rm mrad}\cdot {\rm GeV}/E_{e^\pm}, \label{eq:theta_1e}\\ 
  \theta_2 
  &=
  \left\{ \begin{array}{ll}
    max\qty(\frac{\sqrt{m_{A'}m_e}}{E_e},\qty(\frac{m_{A'}}{E_e})^{\frac{3}{2}}), & :\textrm{bremsstrahlung}\\[1mm]
    0 &: \textrm{annihilation},
  \end{array}\right.\\
  \theta_3 &= \frac{\pi m_{A'}}{2E_{A'}}.
\end{align}
It is required that $r_\perp<r_{det}$, then we obtain the  acceptance as
\begin{equation}
  \mathcal{A}_{e}^{brem,ann} \simeq  e^{-(L_{dump}+L_{sh})/L_{A'}}\qty(e^{-z_{min}/L_{A'}}-e^{-L_{dec}/L_{A'}}),
\end{equation}
where
\begin{equation}
  z_{min}\equiv
  \frac{1}{\theta_3}\qty[(\theta_1+\theta_2)(L_{dump}+L_{sh})+(\theta_1+\theta_2+\theta_3)L_{dec}-r_{det}].
  \label{eq:z_min}
\end{equation}

\item Incoming $\mu^\pm$: Since the flight length of the
  muon is $>\mathcal{O}(10)$ m in the lead shield, the production
  point of the LGB can not be approximated by the front edge of the
  beam dump as the incoming $e^\pm$ case.  We approximate the flight
  length of the muon in matter as
\begin{equation*}
  \delta_\mu = \frac{E_{\mu_0}-E_\mu}{\expval{\dd E/\dd x}},
\end{equation*}
where $E_{\mu_0}$ ($E_\mu$) is the energy of initial (attenuated) muon
in the target. The energy-loss rate in lead is almost energy
independent: $\expval{\dd E/\dd x} \simeq 0.02$ GeV/cm.\footnote
{For the relativistic muon with $E_\mu<1$ TeV, it can be approximated
  as a minimum ionizing particle (mip). T This means that the energy
  loss of mip in the target is approximated by the minimum ionizing
  energy, that is, the energy loss is approximately energy
  independent.}
We neglect the energy loss of the muon in water. Then, we obtain
$r_\perp$ as
\begin{equation}
  r_\perp \simeq \theta_1(L_{dump}+\delta_\mu)+(\theta_1+\theta_2)(L_{shield}+z-\delta_\mu)+(\theta_1+\theta_2+\theta_3)(L_{dec}-z),
\end{equation}
where $\theta_1$ is the muon angle from the beam axis while
$\theta_{2,3}$ are the same as the $e^\pm$ case. The typical value of
$\theta_1$ is
\begin{equation}
  \theta_1 = \sqrt{\qty(\frac{2m_\mu}{E_{\mu_0}})^2+\theta_0},
  \label{eq:theta_1}
\end{equation}
where $\theta_0$ is the standard variance of the angular distribution
of the multiple coulomb scattering (see Ref.\ \cite{Sakaki:2020mqb} for the concrete expression of $\theta_0$).  
Consequently, we obtain
\begin{equation}
  \mathcal{A}^\mu_{e} \simeq e^{-(L_{sh}-\delta_\mu)/L_{A'}}\qty(e^{-z_{min}^\mu/L_{A'}}-e^{-L_{dec}/L_{A'}}),
\end{equation}
where
\begin{equation}
  z_{min}^{\mu} \equiv
  \frac{1}{\theta_3}\qty[-\theta_2\delta_\mu+\theta_1L_{dump}+(\theta_1+\theta_2)L_{shield}+(\theta_1+\theta_2+\theta_3)L_{dec}-r_{det}].
\end{equation}

\end{itemize}

\subsection{Event rate}
The remaining quantities to calculate the event rate in
Eq.\ (\ref{eq:rough_N}) are $N_l$ and $N_{target}$. The number of
lepton in the beam $N_l$ is determined by the experimental setup. For
example, $N_e$ in the ILC experiment is about $4\times10^{21}$ per year
\cite{Behnke:2013xla, Baer:2013cma, Adolphsen:2013jya,
  Adolphsen:2013kya, Behnke:2013lya}. $N_{target}$ is estimated by
using the track length $l_m$, which is the total flight length of the
particle $m$ including the effects from the secondary particles:
\begin{equation}
  N_{target}(E_m) = \sum_{m} \frac{N_{Avo}\rho}{A}\frac{\dd l_m}{\dd E_m}(E_m),
\end{equation}
where $N_{Avo}$ is the Avogadro number and $\rho\ (A)$ is the density (mass number) of the target. 

Based on the discussion so far, we can calculate the event rate. For the event when the LGB is produced by the $e^\pm$ bremsstrahlung process,
\begin{align}
  N_{e}^{brem} = 
  B_{vis}N_e\frac{N_{Avo}\rho}{A}\int_{m_{A'}+m_e}^{E_{beam}}\dd E_e\int_{m_{A'}}^{E_e-m_e}\dd E_{A'}\sum_{e^\pm}\frac{\dd l_{e^\pm}}{\dd E_e}\frac{1}{E_e}\qty[\frac{\dd\sigma}{\dd x}]_{x=\frac{E_{A'}}{E_e}}\mathcal{A}_e^{brem},
  \label{eq:N_ebrem}
\end{align}
where $B_{vis}$ is the branching ratio for the visible particles.\footnote{In Eq.\ (\ref{eq:N_ebrem}), the lower limit of $E_{A'}$ is set to be $m_{A'}$. When $m_{A'}$ is small, $N^{(brem)}$ includes signals with soft final state particles and the event rate is overestimated in this calculation. When the energy cut at the detector is considered, signals with soft particles are rejected and the discovery reach is modified\ \cite{Bauer:2018onh}.}
For the event when the LGB is produced by the annihilation process,
\begin{align}
  N_{e}^{ann} &= N_e\qty(\frac{N_{Avo}\rho}{A}Z)\int_{m_{A'}+m_e}^{E_{beam}}\dd E_e\frac{\dd l_{e^{+}}}{\dd E_e}\sum_{\ell=e,\mu}\sigma_{ann}^\ell\Theta(E_{CM}-2m_\ell)\mathcal{A}_e^{ann}.
\end{align}
For the event when the LGB is produced by the secondary $\mu^\pm$ bremsstrahlung process, \footnote{For the primary muon case, see Sec.\ \ref{sec:muon_bd}.}
\begin{align}
  N_{\mu} = B_{vis}N_e\frac{N_{Avo}\rho}{A}\int_{m_\mu}^{E_{beam}} \dd E_{\mu_0}
  \int_{m_{A'}+m_\mu}^{E_{\mu_0}}&\dd E_\mu\nonumber\\
  \int_{m_{A'}}^{E_\mu-m_e}\dd E_{A'}
  &\frac{\dd l_{\mu}}{\dd E_\mu}\frac{\dd Y_{\mu_0}}{\dd E_{\mu_0}}\frac{1}{E_\mu}\qty[\frac{\dd\sigma}{\dd x}]_{x=\frac{E_{A'}}{E_\mu}}\mathcal{A}^{\mu}_e,
\end{align}
where $Y_{\mu_0}$ is the energy distribution function of muons in the target when the electron is injected \cite{Sakaki:2020cux}:
\begin{equation}
  \frac{\dd Y_{\mu_0}}{\dd E_{\mu_0}} = \frac{0.572E_{beam}}{\ln(183Z^{-1/3})}\qty(\frac{m_e}{m_\mu})^2 \qty(\frac{1}{E_{\mu_0}^2}-\frac{1}{E_{beam}^2}).
\end{equation}

\section{The $e^\pm$ Beam Dump Experiment}
\label{sec:discovery}
\setcounter{equation}{0}

In this section, we show the discovery reach for the LGB in the
$e^\pm$ beam dump experiment.  Before showing the results, we summarize
the concrete experimental setup adopted in this section:
\begin{itemize}
  \item The target is $\rm H_2O$ in the beam dump and the length is
    $L_{dump} = 11\ \rm{m}$ ($30X_0$, with $X_0$ being the radiation
    length).
  \item The shield is installed behind the beam dump.  We assume zero
    background due to the shield.  The shield length is assumed as
    $L_{shield}=50\ \rm{m}$.
  \item The decay volume is installed behind the shield. The length of the decay volume is assumed as $L_{dec}=50\ \rm{m}$.
  \item The detector is installed behind the decay volume. The radius of the detector is assumed as $r_{det}=2\ \rm{m}$.

  \item The ILC bunch train contains 1312 bunches and each bunch has
    $2\times 10^{10}\ e^\pm$. The frequency of the dump of the bunch
    train is $5\ \rm{Hz}$. Thus $4\times 10^{21}\ e^\pm$ are injected
    with a 1-year operation \cite{Behnke:2013xla, Baer:2013cma,
      Adolphsen:2013jya, Adolphsen:2013kya, Behnke:2013lya}.
  \item We consider the ILC-250, 500, 1000, corresponding to the beam
    energy $E_{beam}=125,\ 250,\ 500\ \rm{GeV}$.
\end{itemize}
Notice that the shield length directly affects the sensitivity. With longer shield, the sensitivity to the
short-lived LGB becomes worse.

\subsection{$U(1)_{L_e-L_\mu}$, $U(1)_{L_e-L_\tau}$ Models}

\begin{figure}[t]
  \centering
  \begin{tabular}{cc}
  
  \begin{minipage}{0.5\hsize}
      \centering
      \includegraphics[width=1.0\linewidth]{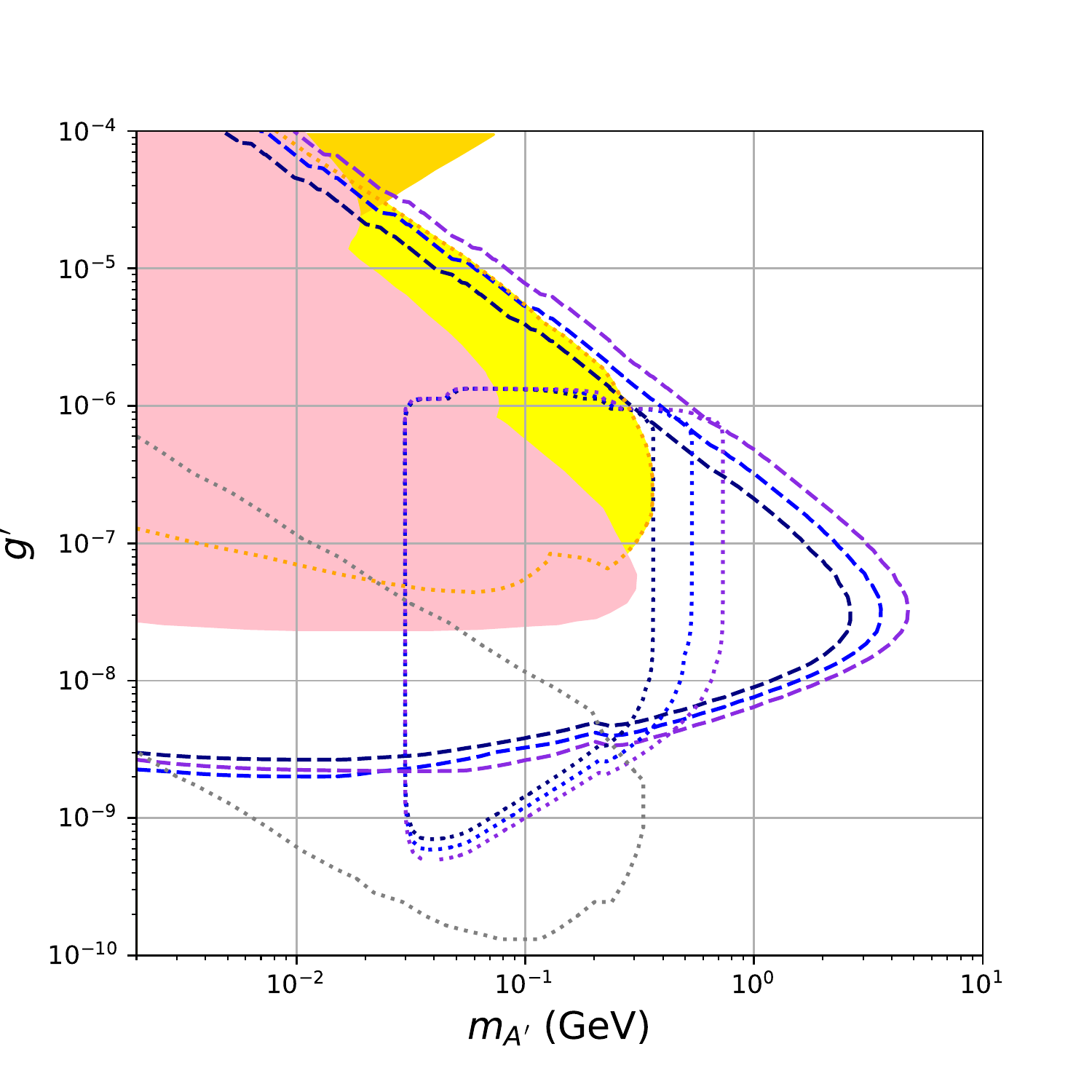}
      \subcaption{electron beam}
  \end{minipage}
  
  \begin{minipage}{0.5\hsize}
      \centering
      \includegraphics[width=1.0\linewidth]{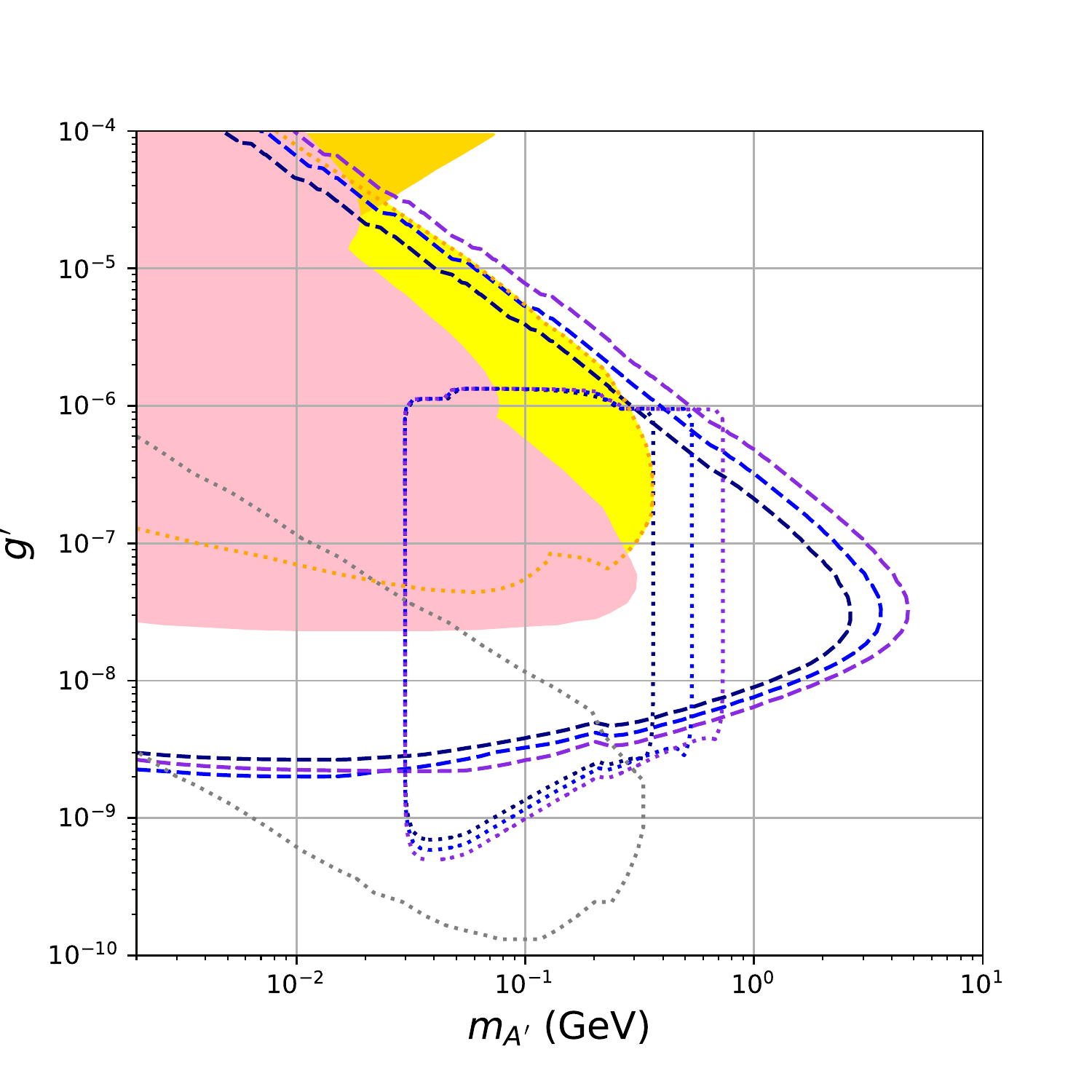}
      \subcaption{positron beam}
  \end{minipage}
  
  \end{tabular}
  \caption{The ILC beam dump experiment discovery reach for $10$ years
    operation for the $U(1)_{L_e-L_\mu}$ model.  Left (right) figure
    shows the result with electron (positron) beam. Line colors
    correspond to the beam energy: 125 GeV (navy), 250 GeV (blue) and
    500 GeV (purple). Line styles correspond to the production
    process: bremsstrahlung (dashed line) and annihilation (dotted line).  The
    pink-shaded region is excluded by the previous beam dump
    experiments \cite{Bauer:2018onh}. The orange-shaded region is
    excluded by the electron-neutrino scattering experiment, Texono
    \cite{TEXONO:2009knm}. The grey-shaded region is excluded by the
    observation of the supernova \cite{Escudero:2019gzq}.  The
    yellow-shaded region is the expected sensitivity of the SHiP
    experiment \cite{Alekhin:2015byh}. }
  \label{fig:ILC_emu}

\end{figure}

\begin{figure}[t]
  \begin{tabular}{cc}
    
    \begin{minipage}{0.5\hsize}
      \centering
      \includegraphics[width=1.0\linewidth]{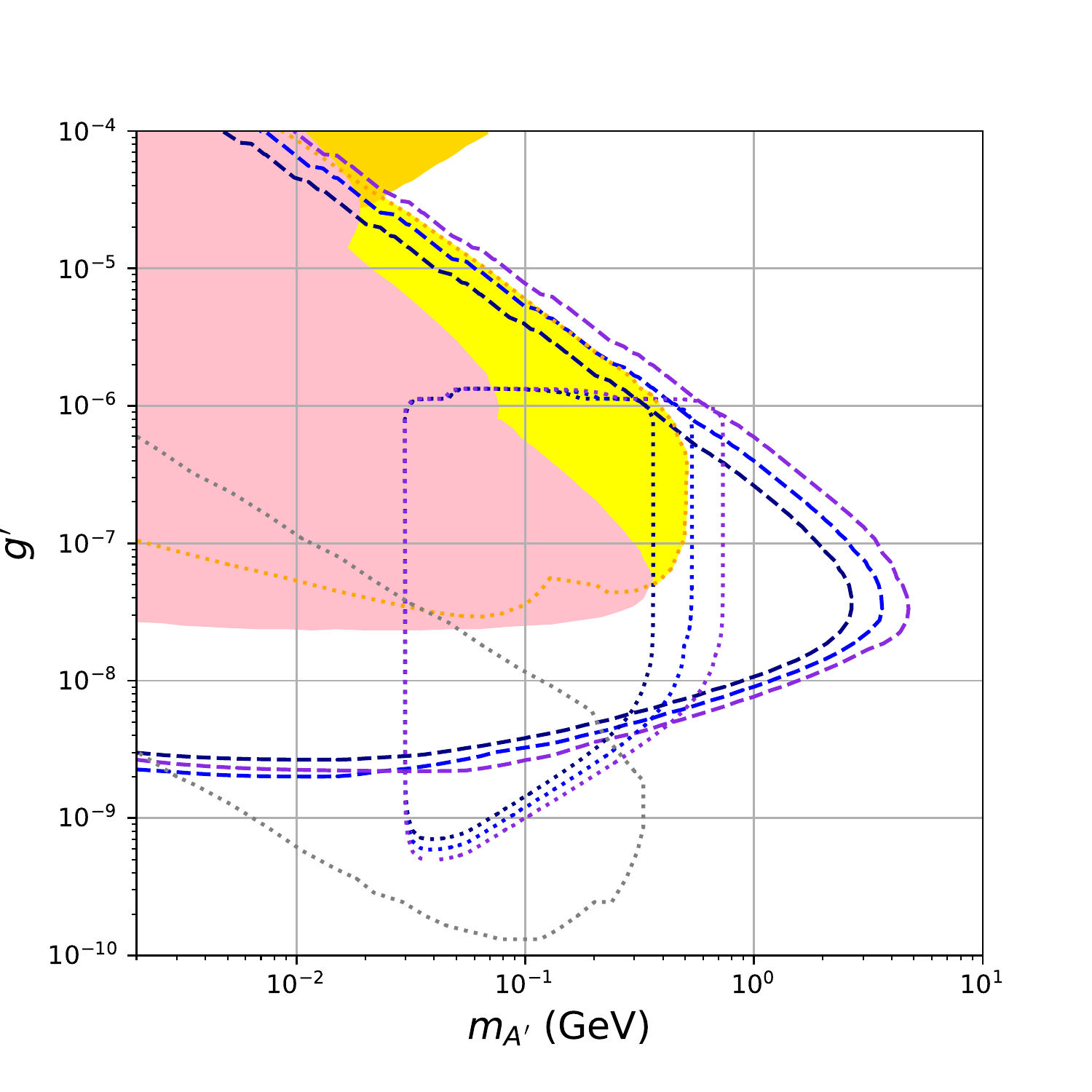}
      \subcaption{electron beam}
    \end{minipage}
    
    \begin{minipage}{0.5\hsize}
      \centering
      \includegraphics[width=1.0\linewidth]{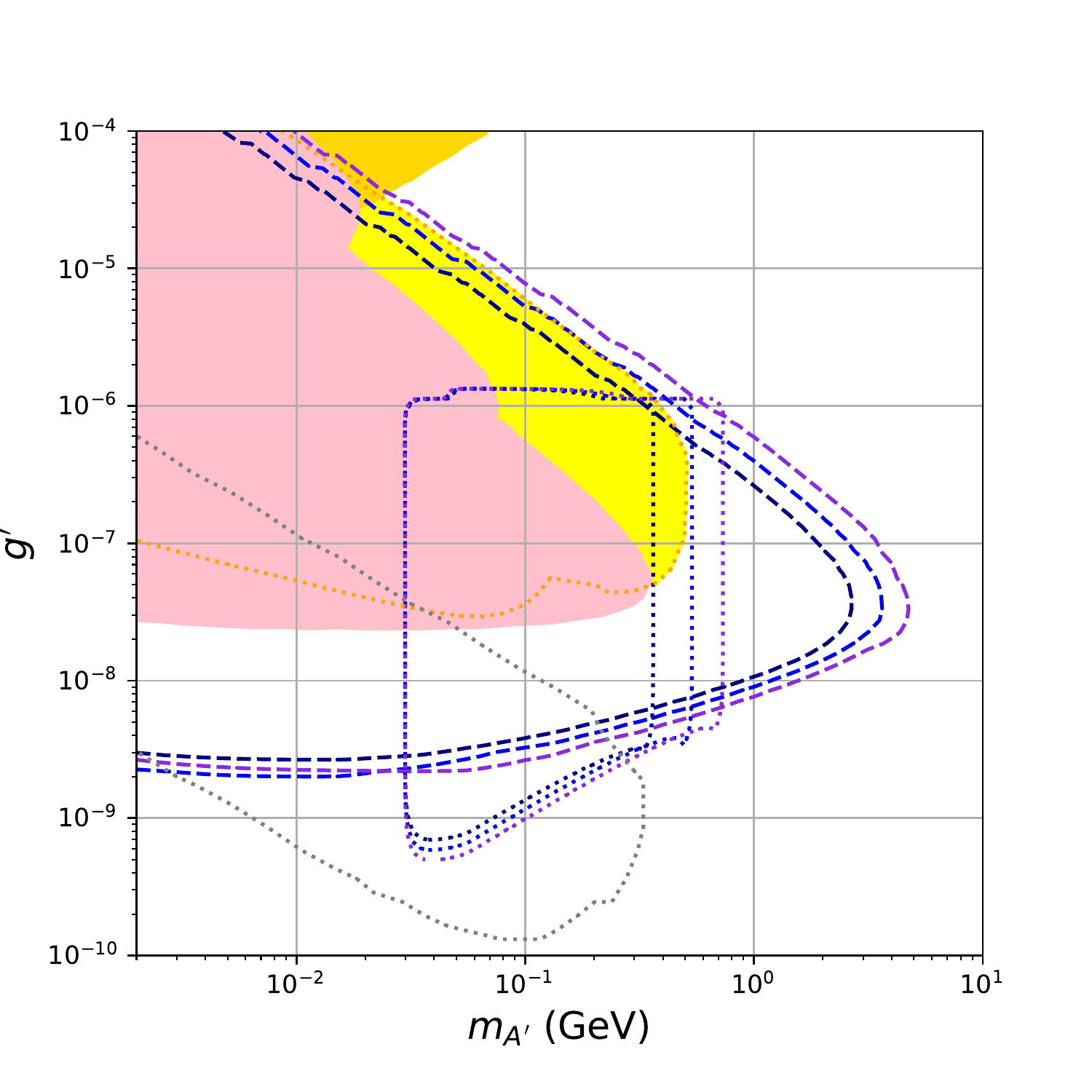}
      \subcaption{positron beam}
    \end{minipage}
    
  \end{tabular}
  \caption{Same to Fig. \ref{fig:ILC_emu}, but for the
    $U(1)_{L_e-L_\tau}$ model.}
  \label{fig:ILC_etau}
\end{figure}

\begin{figure}[ht]
  \centering
  \begin{tabular}{ll}
  
    \begin{minipage}{0.5\hsize}
        \centering
        \includegraphics[width=1.0\linewidth]{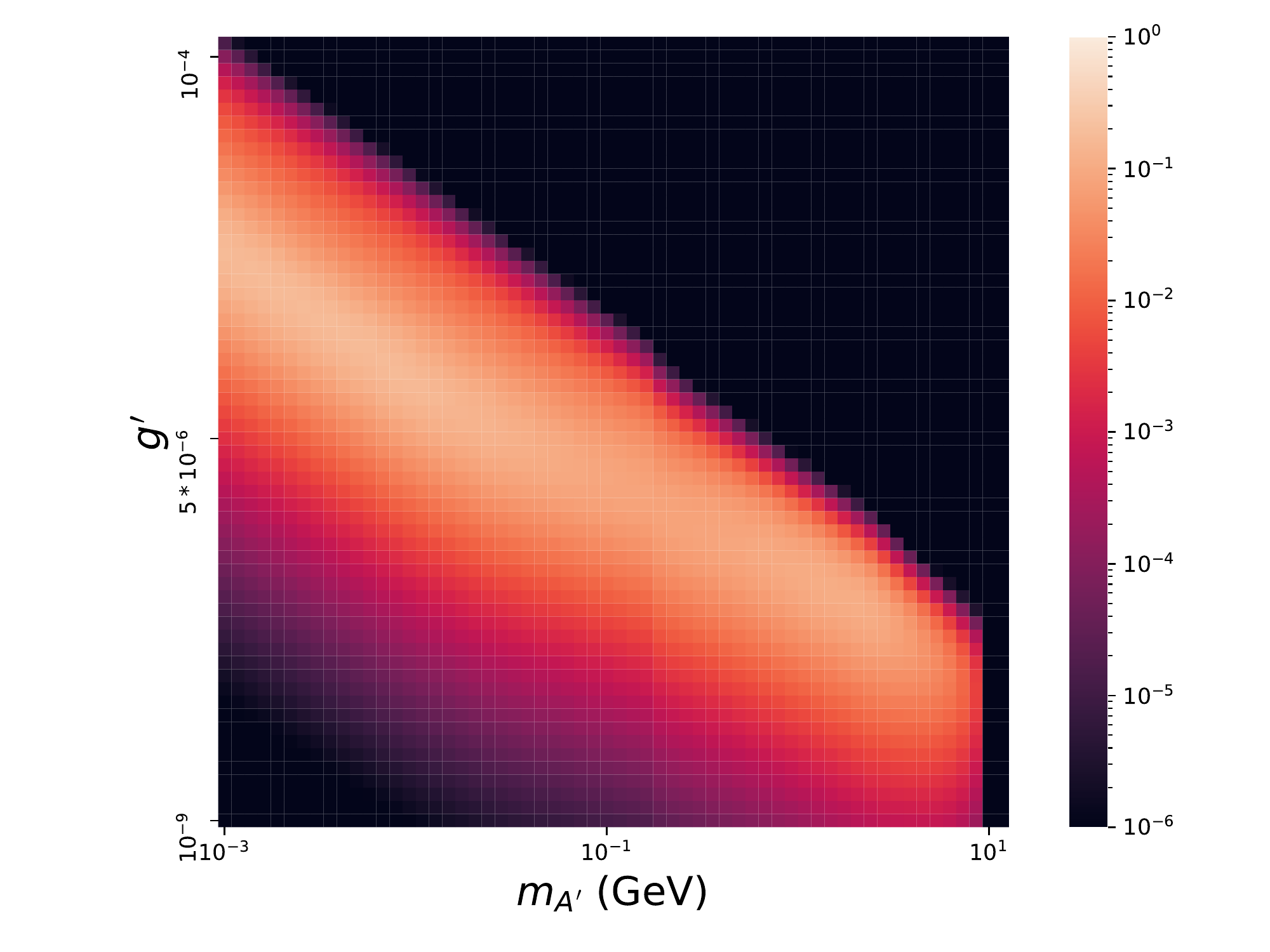}
        \subcaption{Bremsstrahlung}
    \end{minipage}
    \begin{minipage}{0.5\hsize}
        \centering
        \includegraphics[width=1.0\linewidth]{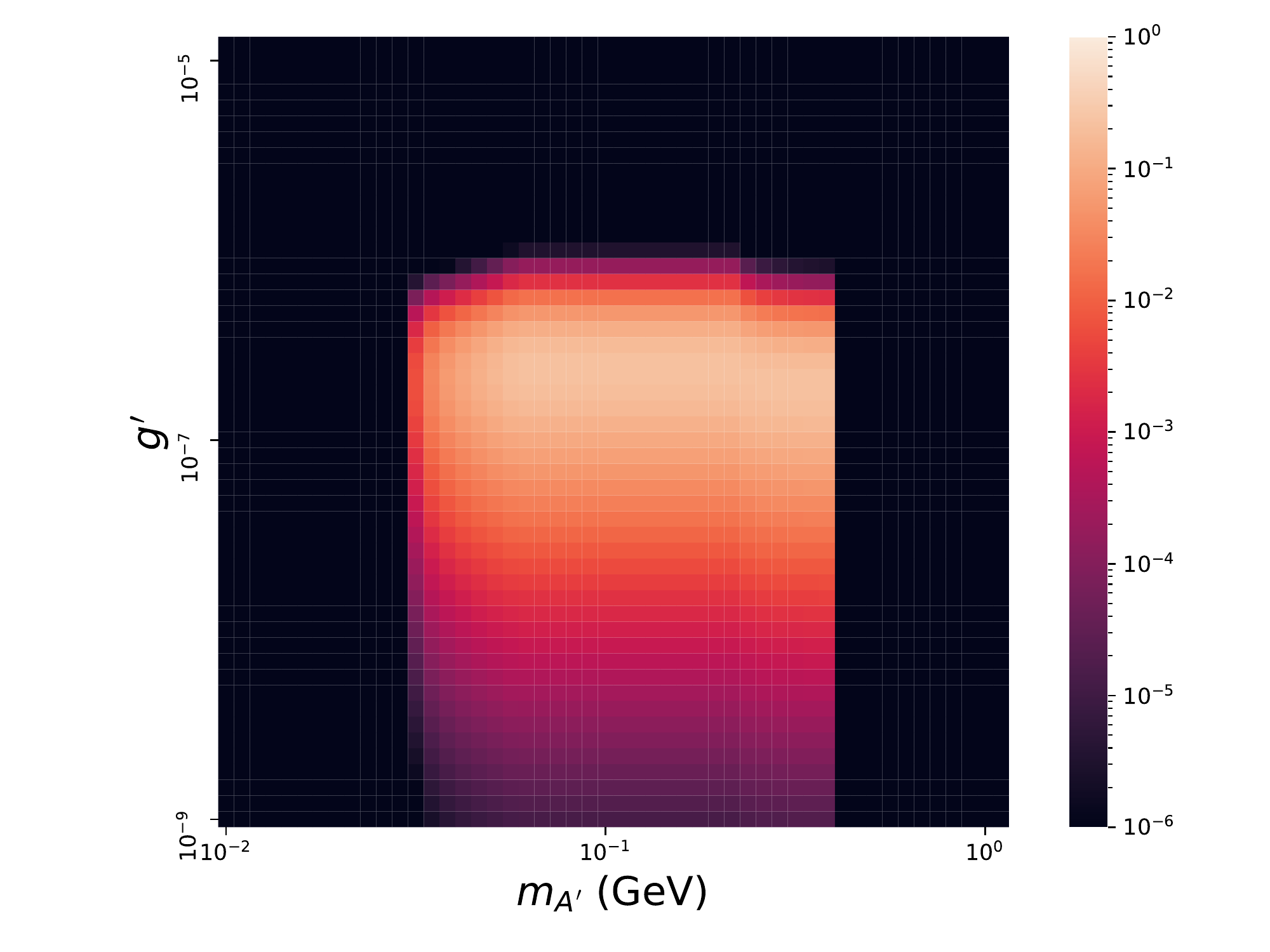}
        \subcaption{Annihilation}
    \end{minipage}
    
    \end{tabular}
    \caption{Heatmap of the acceptance ($i.e.$, the ratio of the
      number of LGBs produced in the dump to the number of signals) on
      $m_{A'}$-$g'$ plane. In this figure, we consider the LGB associated with $U(1)_{L_e-L_\mu}$, but the behavior is not so different when we consider the other models.}
    \label{fig:acceptance}
\end{figure}

In Figs.\ \ref{fig:ILC_emu} and \ref{fig:ILC_etau}, we show the
contours of the number of signal events being equal to 3, which we
call the discovery reaches.  In
Fig.\ \ref{fig:acceptance}, we also show the acceptance ($i.e.$, the
ratio of the number of LGBs produced in the dump to the number of
signals) of the $U(1)_{L_e-L_\mu}$ model on $m_{A'}$-$g'$ plane for
the bremsstrahlung and annihilation processes.
In Figs.\ \ref{fig:ILC_emu} and \ref{fig:ILC_etau}, the yellow-shaded region is the expected sensitivity
of the SHiP experiment, a future proton beam dump experiment
\cite{Alekhin:2015byh}. For the LGB search, the $e^\pm$ beam dump
experiment has an advantage compared with the proton beam dump
experiment. Since the LGBs of $U(1)_{L_e-L_\mu}$ and  $U(1)_{L_e-L_\tau}$
models couple to the quarks only via the loop-suppressed kinetic
mixing, the sensitivity of hadronic beam dump experiments is worse than the $e^\pm$ case. This is the reason why the ILC beam dump experiment has better
sensitivity than the SHiP experiment.  We can see that the ILC beam dump
experiment has a sensitivity to the unexplored region. 

We comment on the behavior of the discovery reach.  The
sensitivity becomes worse as the coupling constant $g'$ becomes too
large or too small.  In the large $g'$ case, the LGBs are abundantly
produced but most of them decays before reaching to the decay volume.
The behavior of the discovery reach in the large coupling region can be understood from
the fact that the decay probability depends exponentially on the
combination $\frac{m_{A'}^2g'^2}{E_{A'}}$, with $E_{A'}$ being the
energy of the LGB.  In the bremsstrahlung process, $E_{A'}$ does not
depend on $m_{A'}$ and $g'$, so the upper edge is along
$m_{A'}g'\sim const.$. In the annihilation process,
$E_{A'}=\frac{m_{A'}^2}{2m_e}$ and contour is along $g'\sim const.$.  In
the small $g'$ region, the production rate of LGB in the dump and the
decay probability in the decay volume are both suppressed.  The number
of events is proportional to $g'^4$ in the bremsstrahlung process and
$\frac{g'^4}{m_{A'}^2}$ in the annihilation process.\footnote
{In our calculation, we set the lower limit of $E_{A'}$ integral by
  $m_{A'}$, instead of $E_{cut}$. This induces the $m_{A'}$ dependence
  to the bremsstrahlung sensitivity
  \cite{Bauer:2018onh,niki:2022m}. Moreover, the number of produced
  LGB from the annihilation process depends on $E_e$ because the track
  length depends on $E_e$ \cite{Asai:2021ehn}. This behavior is valid
  when $E_{e}\sim E_{beam}$.}
We also comment that the sensitivity region of the annihilation
process is sharply terminated at high and low values of the LGB mass.
In the annihilation process, $\sqrt{2m_e E_{e^+}}=m_{A'}$ holds under
the narrow-width approximation. Then, $E_{e^+}<E_{beam}$ sets an upper bound on $m_{A'}$ which is kinematically accessible with the annihilation
process.
For the ILC-250, for example, the upper bound is about $300$ MeV. This
limit makes the right edge of the sensitivity region from the
annihilation production.  The left edge is from the LGB-mass
dependence of $E_{A'}$.  As we consider smaller $m_{A'}$, the energy
of the LGB produced by the annihilation process becomes lower.
Because the angle $\theta_1$ of the positron becomes larger as the
positron energy becomes smaller (see Eq.\ (\ref{eq:theta_1e})), the
annihilation process loses the sensitivity in the low mass region.
This sets the left edge of the discovery reach from the annihilation
process.

Comparing Figs.\ \ref{fig:ILC_emu} and \ref{fig:ILC_etau}, the
discovery reaches for the $U(1)_{L_e-L_\mu}$ and $U(1)_{L_e-L_\tau}$
models are similar.  This is because the LGBs of both models directly
couple to the electron.  On the other hand, the dominant (visible)
decay modes of the LGB are different.  For the case of
$U(1)_{L_e-L_\mu}$, the dominant visible decay modes are
$A'\rightarrow e^+e^-$ and $\mu^+\mu^-$, while the LGB of the
$U(1)_{e-\tau}$ model decays mostly into $e^+e^-$ (if $m_X<2m_\tau$).
These features may be used to distinguish the models behind the LGB,
as discussed in Ref.\ \cite{Asai:2021xtg}.

\subsection{$U(1)_{L_\mu-L_\tau}$ Model}

\begin{figure}[t]
  \centering
  \begin{tabular}{cc}
  
  \begin{minipage}{0.5\hsize}
      \centering
      \includegraphics[width=1.0\linewidth]{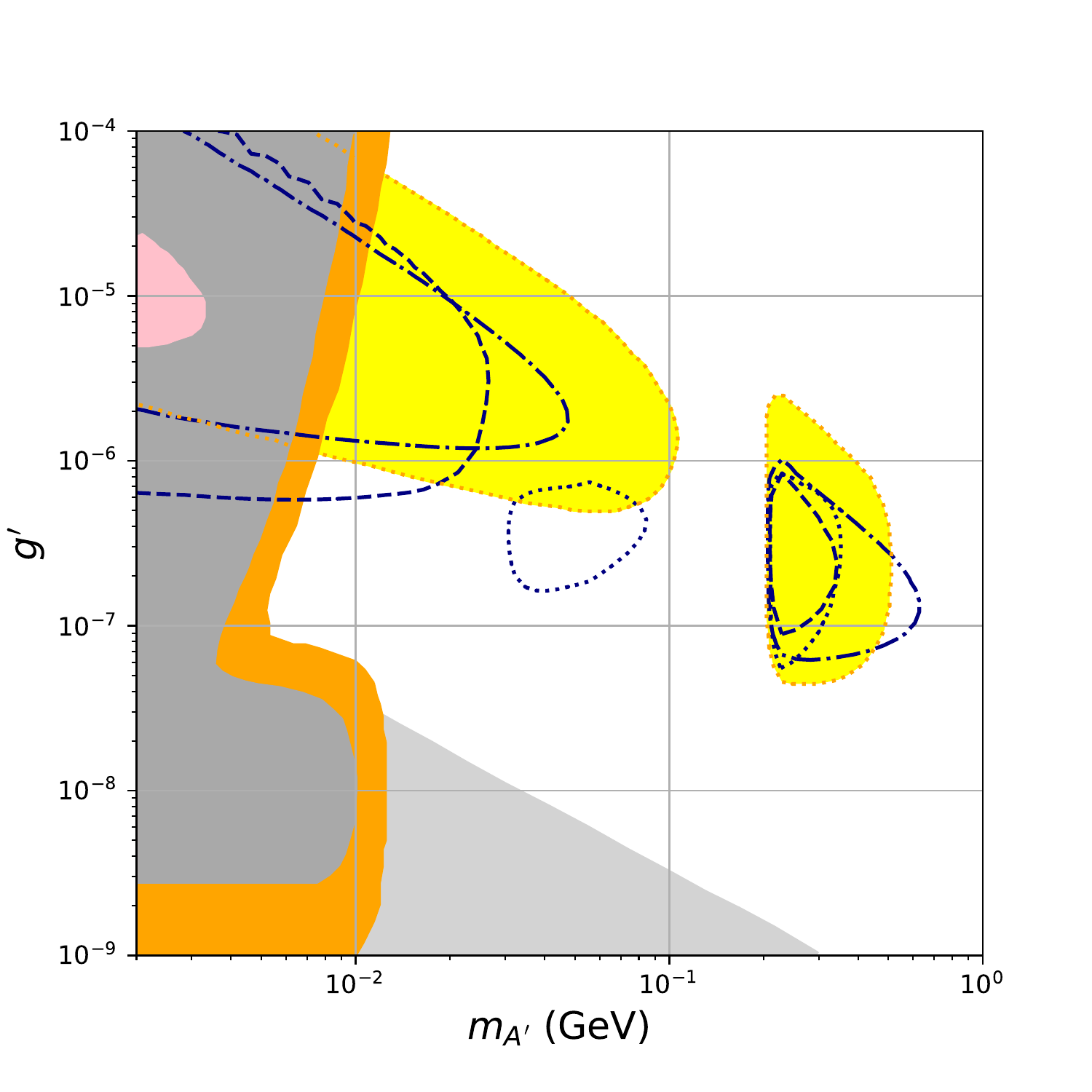}
      \hspace{2cm}
      \subcaption{electron beam}
  \end{minipage}
  
  \begin{minipage}{0.5\hsize}
      \centering
      \includegraphics[width=1.0\linewidth]{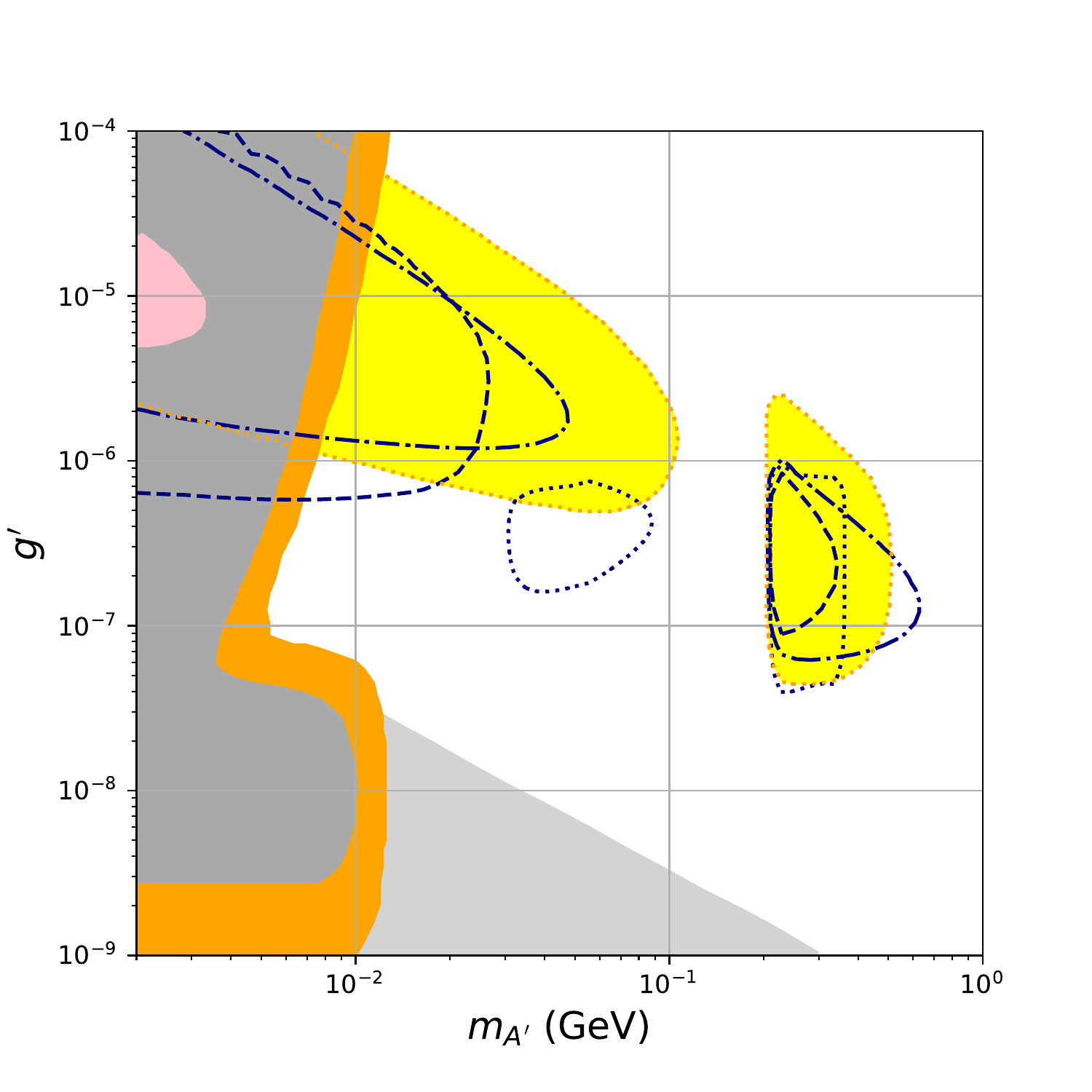}
      \hspace{2cm}
      \subcaption{positron beam}
  \end{minipage}
  
  \end{tabular}
  \caption{The $e^\pm$ beam dump experiment discovery reach for the
    $U(1)_{L_\mu-L_\tau}$ extension model. Several production
    processes of the LGB are considered: $e$ bremsstrahlung (dashed
    line), $\mu$ bremsstrahlung (dash-dotted line), and the
    annihilation (dotted line).  The pink shaded region is excluded by
    the previous beam dump experiment, E137 at SLAC
    \cite{Bjorken:1988as,Bauer:2018onh}. The dark gray region is
    excluded ny the BBN constraint \cite{Kamada:2015era}. The light
    grey shaded region is excluded by the observation of SN1987A
    \cite{Escudero:2019gzq}.  In the orange-shaded region, the Hubble
    tention is alleviated \cite{Escudero:2019gzq}.}
  \label{fig:ILC_mut_ordinary}
\end{figure}

Fig.\ \ref{fig:ILC_mut_ordinary} shows the discovery reach for the LGB
in the $U(1)_{L_\mu-L_\tau}$ model in the ILC-250 beam dump experiment
with a 10-year operation. In addition to the $e^\pm$ bremsstrahlung
and the annihilation production processes, we also consider the muon
bremsstrahlung production process.\footnote
{For the $U(1)_{L_e-L_\mu}$ model, muon bremsstrahlung process
  contributes to the sensitivity without loop-suppression. Electron
  bremsstrahlung process, however, also contributes to the sensitivity
  without loop-suppression in this model. Since the number of initial
  electrons is much larger than the number of initial muons in the
  electron beam dump experiment, the contributions from the muon
  bremsstrahlung process are expected to be much smaller than the
  contributions from electrons.  The yellow regions show the expected
  sensitivity of the SHiP experiment \cite{Alekhin:2015byh}.}
The muon bremsstrahlung process is important for the
$U(1)_{L_\mu-L_\tau}$ model because the LGB directly couples to muon.
We can see that the ILC beam dump experiment has a sensitivity to the
region which has not been excluded yet.
Particularly, the LGB in the
$U(1)_{L_\mu-L_\tau}$ model is motivated by the fact that it may
alleviate the Hubble tension \cite{Escudero:2019gzq}. The ILC beam
dump experiment can accesss the part of the region suggested by the
Hubble tension.

In the previous study of the search of the LGBs with the ILC beam dump
experiment \cite{Asai:2021xtg}, only the production process by the
beam particle was considered. We can see that the effects of the
secondary particles extend the sensitivity to the small-coupling region and
that the muon bremsstrahlung effects extend the sensitivity to the
large-mass region. Due to the large cross section and the different
kinematics, we can also see that the signal produced by the annihilation
process may cover the parameter region uncovered by the bremsstrahlung
process.  The annihilation process requires the 
positron in the initial state. For better sensitivity from the annihilation production, the
use of a positron beam is better.

\section{The Muon Beam Dump Experiment}
\label{sec:muon_bd}
\setcounter{equation}{0}

\begin{figure}[t]
  \centering
  \begin{tabular}{cc}
  
  \begin{minipage}{0.5\hsize}
      \centering
      \includegraphics[width=1.0\linewidth]{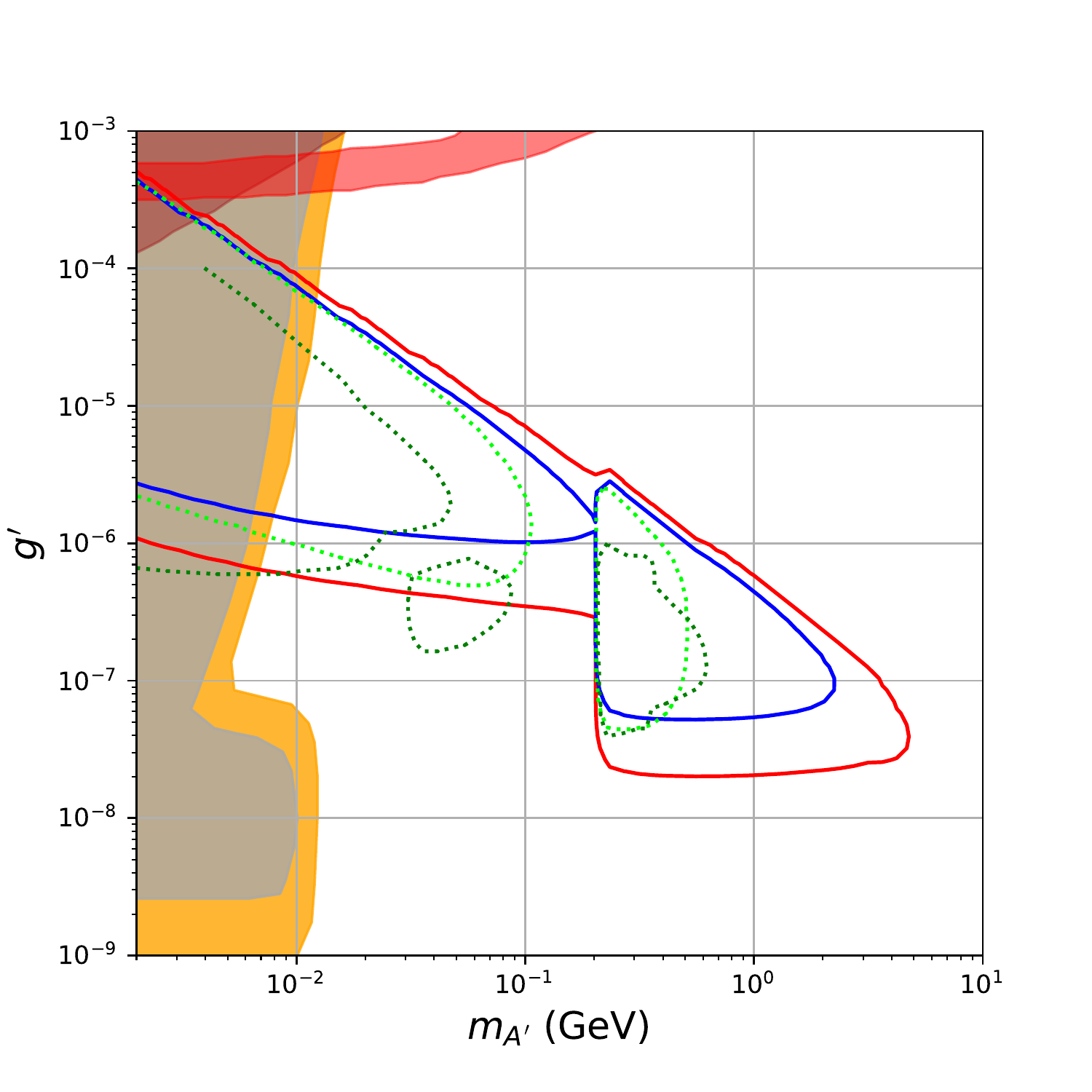}
      \hspace{2cm}
  \end{minipage}
  
  \begin{minipage}{0.5\hsize}
      \centering
      \includegraphics[width=1.0\linewidth]{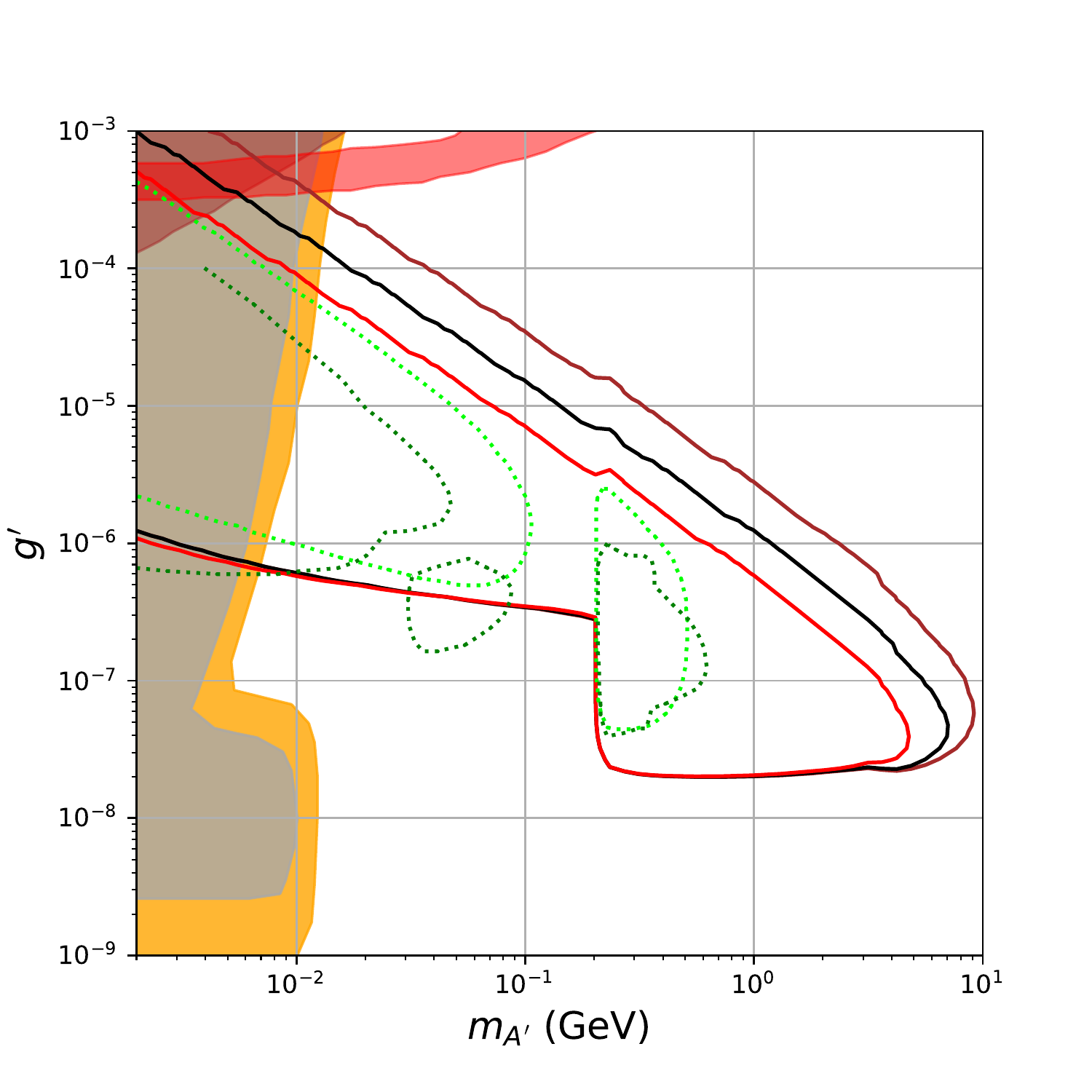}
      \hspace{2cm}
  \end{minipage}
  
  \end{tabular}
  \caption{The muon beam dump experiment discovery reach for the
    case of $U(1)_{L_\mu-L_\tau}$ extension model (solid lines).  The
    grey-shaded (brown-shaded) region is excluded by the BBN
    observation \cite{Escudero:2019gzq} (the solar neutrino
    observation \cite{Bellini:2011rx, Kaneta:2016uyt}).  In the
    orange-shaded region, the Hubble tension is alleviated by the LGB
    \cite{Escudero:2019gzq}.  The red-shaded region is the one motivated by the
    muon $g-2$ anomaly.  The green (light green) dotted line shows the
    expected discovery reach in the ILC beam dump experiment (SHiP
    experiment), as shown in Sec.\ \ref{sec:discovery}.  The left
    figure shows the case with $200$ m shield and lead (red) or water
    (blue) target. The right figure shows the case with lead target
    and several length shield: 200 m (red), 50 m (brown), and 10 m
    (black).}
  \label{fig:muon_bd}
\end{figure}

In this section, we consider the beam dump experiment using a muon
beam at the muon collider.\footnote
{Since muons are not stopped by the target, the target is not a
  beam dump. We, however, call the experiment as ``beam-dump experiment''
as we did in the case of $e^+e^-$ colliders.}
The muon beam dump experiment has advantages to search for muon-philic
new particles.  In particular, in the $U(1)_{L_\mu-L_\tau}$ model, the
LGB does not couple to electron at tree level and then the production
cross section is suppressed in the $e^\pm$ beam dump experiment.  We
expect that the muon beam dump experiment has a good sensitivity to
the LGB in the $U(1)_{L_\mu-L_\tau}$ model.  Thus, in this section, we
study the search of the LGB in the $U(1)_{L_\mu-L_\tau}$ model at
the muon beam dump experiment.  Such a subject was already
considered in Ref.\ \cite{Cesarotti:2022ttv}.  In the paper, however, the
effect of the loop-induced kinetic mixing was not taken into account,
so the sensitivity below the di-muon mass was underestimated.  In the
present study, the loop-induced kinetic mixings is included in
deriving the discovery reach.

We consider the experimental setup similar to the one in the $e^\pm$
beam dump experiment.  We consider a lead target as well as a water
target for comparison; the length of the target is taken to be
$11\ {\rm m}$.  Then, the shield and the decay volume with the
detectors are assumed to be installed behind the target.  As a source
of the muon beam, we consider the US Muon Accelerator Program (MAP)
design, where the number of muons in the beam is $N_\mu\sim 10^{20}$
per year and the beam energy is $1.5$ TeV \cite{Delahaye:2013jla}.  In
the muon beam dump experiment based on the MAP, the muons in the beam are
very energetic and we may have serious muon background.  In the SHiP
experiment, for example, the muons with the energy $\sim 400$ GeV are
expected to be removed by $50\ \rm{m}$ active shield.  With rescaling
this result, $1.5$ TeV muons may be removed by using $\sim 200$ m
active shield; we take $L_{shield}=200\ {\rm m}$ as our canonical
value of the shield length.  We also consider the cases with shorter
shield length which may be realized if a stronger magnet or additional
technical progresses are developed in the future.

As the event rate in the $e^\pm$ beam dump experiment, the event rate in the muon beam dump experiment can be estimated as
\begin{align}
  N = B_{vis}N_\mu\frac{N_{Avo}\rho}{A}\int_{m_\mu+m_{A'}}^{E_{beam}} \dd E_{\mu}
  \int_{m_{A'}}^{E_\mu-m_e}\dd E_{A'}
  \frac{\dd l_{\mu}}{\dd E_\mu}\frac{1}{E_\mu}\qty[\frac{\dd\sigma}{\dd x}]_{x=\frac{E_{A'}}{E_\mu}}\mathcal{A}_\mu
  \Theta(L_{dump}-\delta_\mu).
  \label{eq:event_mubd}
\end{align}
For the calculation of the acceptance $\mathcal{A}_\mu$, we take
$\theta_1=\theta_{0}$ (see Eq.\ (\ref{eq:theta_1})) and we obtain
\begin{equation}
  \mathcal{A}_\mu \simeq e^{-(L_{dump}+L_{shield}-\delta_\mu)/L_{A'}}\qty(e^{-z_{min}/L_{A'}}-e^{-L_{dec}/L_{A'}}),
\end{equation}
where
\begin{equation*}
  z_{min} = \frac{1}{\theta_3}\qty[-\theta_2\delta_\mu+(\theta_1+\theta_2)(L_{dump}+L_{shield})+(\theta_1+\theta_2+\theta_3)L_{dec}-r_{det}].
\end{equation*}
The theta function in Eq.\ (\ref{eq:event_mubd}) takes care of the
fact that the 
muons do not stop in the dump.\footnote
{In this setup, we assume the shield is active, that is, muons will be
  removed by the magnetic field. Then the muon does not produce the
  LGB in the shield.}
Notice that 
$\theta_1$ and $z_{min}$ are calculated by using the
fact that the initial-state muon is the beam
particle not the secondary one.

Fig.\ \ref{fig:muon_bd} shows the discovery reach for the LGB in the
$U(1)_{L_\mu-L_\tau}$ model with the muon beam dump experiment based
on the MAP.  For the case of the $U(1)_{L_\mu-L_\tau}$ model, we can
see that the muon beam dump experiment has better sensitivity than the
future electron and proton beam dump experiments, especially for the
case of large LGB mass.  We can see that the muon beam dump experiment
can access the parameter region suggested by the Hubble tension.
Unfortunately, the muon beam dump experiment hardly reach the region
explaining the muon $g-2$ anomaly.

\section{Summary and Discussion}
\label{sec:summary}
\setcounter{equation}{0}

In this paper, we have discussed the possibility to search for the LGBs
in the lepton beam dump experiments. In the $e^\pm$ beam dump
experiment, we calculate the event rate taking into account the
effects of the secondary particles which were not considered in the
previous study
\cite{Asai:2021xtg}. Then we consider the production processes not
only the electron bremsstrahlung but also the annihilation process and
the muon bremsstrahlung process.  Since the annihilation process has
fewer vertices than the bremsstrahlung process, the cross section
tends to be larger
than that of the bremsstrahlung process. We have shown that
these new production processes extend the discovery reaches for the
LGB (see Fig. \ref{fig:ILC_emu} -- \ref{fig:muon_bd}).  We also
considered the search for the LGB of the $U(1)_{L_\mu-L_\tau}$ model
in the muon beam dump experiment. We include the effect of the
loop-induced kinetic mixing, which was not considered in the previous
work \cite{Cesarotti:2022ttv}.
 For muon-philic particles, in particular, the LGB
of the $U(1)_{L_\mu-L_\tau}$ model, the muon beam dump experiment
gives a better sensitivity than the electron ones. 

There are the other LGB production processes, for example, the decay of 
hadrons. Hadrons can be produced in the lepton beam dump via the
photoproduction processes \cite{Schuster:2021mlr} and then the some of
hadrons decay into the LGB. This process may extend the sensitivity as
in the SHiP experiment. Such an issue will be discussed in elsewhere.

\vspace{3mm}
\noindent{\it Acknowledgments:} 
The work was supported
by JSPS KAKENHI Grant No.\ 16H06490 (TM), 18K03608 (TM) and 22H01215 (TM), 22J21016 (AN).


\bibliographystyle{jhep}
\bibliography{ref}

\providecommand{\href}[2]{#2}\begingroup\raggedright\begin{thebibliography}{10}

\bibitem{Behnke:2013xla}
T.~Behnke, J.E.~Brau, B.~Foster, J.~Fuster, M.~Harrison, J.M.~Paterson et~al.,
  eds., \emph{{The International Linear Collider Technical Design Report -
  Volume 1: Executive Summary}},
  \href{https://arxiv.org/abs/1306.6327}{{\ttfamily 1306.6327}}.

\bibitem{Baer:2013cma}
H.~Baer et~al., eds., \emph{{The International Linear Collider Technical Design
  Report - Volume 2: Physics}},
  \href{https://arxiv.org/abs/1306.6352}{{\ttfamily 1306.6352}}.

\bibitem{Adolphsen:2013jya}
C.~Adolphsen et~al., eds., \emph{{The International Linear Collider Technical
  Design Report - Volume 3.I: Accelerator \textbackslash{}\& in the Technical
  Design Phase}},  \href{https://arxiv.org/abs/1306.6353}{{\ttfamily
  1306.6353}}.

\bibitem{Adolphsen:2013kya}
C.~Adolphsen et~al., eds., \emph{{The International Linear Collider Technical
  Design Report - Volume 3.II: Accelerator Baseline Design}},
  \href{https://arxiv.org/abs/1306.6328}{{\ttfamily 1306.6328}}.

\bibitem{Behnke:2013lya}
H.~Abramowicz et~al., \emph{{The International Linear Collider Technical Design
  Report - Volume 4: Detectors}},
  \href{https://arxiv.org/abs/1306.6329}{{\ttfamily 1306.6329}}.

\bibitem{ILCInternationalDevelopmentTeam:2022izu}
{\scshape ILC International Development Team} collaboration, \emph{{The
  International Linear Collider: Report to Snowmass 2021}},
  \href{https://arxiv.org/abs/2203.07622}{{\ttfamily 2203.07622}}.

\bibitem{Roloff:2018dqu}
{\scshape CLIC, CLICdp} collaboration, P.~Roloff, R.~Franceschini, U.~Schnoor
  and A.~Wulzer, eds., \emph{{The Compact Linear e$^+$e$^-$ Collider (CLIC):
  Physics Potential}},  \href{https://arxiv.org/abs/1812.07986}{{\ttfamily
  1812.07986}}.

\bibitem{Delahaye:2019omf}
J.P.~Delahaye, M.~Diemoz, K.~Long, B.~Mansouli\'e, N.~Pastrone, L.~Rivkin
  et~al., \emph{{Muon Colliders}},
  \href{https://arxiv.org/abs/1901.06150}{{\ttfamily 1901.06150}}.

\bibitem{Kanemura:2015cxa}
S.~Kanemura, T.~Moroi and T.~Tanabe, \emph{{Beam dump experiment at future
  electron\textendash{}positron colliders}},
  \href{https://doi.org/10.1016/j.physletb.2015.10.002}{\emph{Phys. Lett. B}
  {\bfseries 751} (2015) 25}
  [\href{https://arxiv.org/abs/1507.02809}{{\ttfamily 1507.02809}}].

\bibitem{Cesarotti:2022ttv}
C.~Cesarotti, S.~Homiller, R.K.~Mishra and M.~Reece, \emph{{Probing New Gauge
  Forces with a High-Energy Muon Beam Dump}},
  \href{https://arxiv.org/abs/2202.12302}{{\ttfamily 2202.12302}}.

\bibitem{Sakaki:2020mqb}
Y.~Sakaki and D.~Ueda, \emph{{Searching for new light particles at the
  international linear collider main beam dump}},
  \href{https://doi.org/10.1103/PhysRevD.103.035024}{\emph{Phys. Rev. D}
  {\bfseries 103} (2021) 035024}
  [\href{https://arxiv.org/abs/2009.13790}{{\ttfamily 2009.13790}}].

\bibitem{Asai:2021ehn}
K.~Asai, S.~Iwamoto, Y.~Sakaki and D.~Ueda, \emph{{New physics searches at the
  ILC positron and electron beam dumps}},
  \href{https://arxiv.org/abs/2105.13768}{{\ttfamily 2105.13768}}.

\bibitem{Asai:2021xtg}
K.~Asai, T.~Moroi and A.~Niki, \emph{{Leptophilic Gauge Bosons at ILC Beam Dump
  Experiment}},
  \href{https://doi.org/10.1016/j.physletb.2021.136374}{\emph{Phys. Lett. B}
  {\bfseries 818} (2021) 136374}
  [\href{https://arxiv.org/abs/2104.00888}{{\ttfamily 2104.00888}}].

\bibitem{Foot:1990mn}
R.~Foot, \emph{{New Physics From Electric Charge Quantization?}},
  \href{https://doi.org/10.1142/S0217732391000543}{\emph{Mod. Phys. Lett. A}
  {\bfseries 6} (1991) 527}.

\bibitem{He:1990pn}
X.G.~He, G.C.~Joshi, H.~Lew and R.R.~Volkas, \emph{{NEW Z-prime
  PHENOMENOLOGY}}, \href{https://doi.org/10.1103/PhysRevD.43.R22}{\emph{Phys.
  Rev. D} {\bfseries 43} (1991) 22}.

\bibitem{He:1991qd}
X.-G.~He, G.C.~Joshi, H.~Lew and R.R.~Volkas, \emph{{Simplest Z-prime model}},
  \href{https://doi.org/10.1103/PhysRevD.44.2118}{\emph{Phys. Rev. D}
  {\bfseries 44} (1991) 2118}.

\bibitem{Foot:1994vd}
R.~Foot, X.G.~He, H.~Lew and R.R.~Volkas, \emph{{Model for a light Z-prime
  boson}}, \href{https://doi.org/10.1103/PhysRevD.50.4571}{\emph{Phys. Rev. D}
  {\bfseries 50} (1994) 4571}
  [\href{https://arxiv.org/abs/hep-ph/9401250}{{\ttfamily hep-ph/9401250}}].

\bibitem{Araki:2012ip}
T.~Araki, J.~Heeck and J.~Kubo, \emph{{Vanishing Minors in the Neutrino Mass
  Matrix from Abelian Gauge Symmetries}},
  \href{https://doi.org/10.1007/JHEP07(2012)083}{\emph{JHEP} {\bfseries 07}
  (2012) 083} [\href{https://arxiv.org/abs/1203.4951}{{\ttfamily 1203.4951}}].

\bibitem{Heeck:2014sna}
J.~Heeck, \emph{{Neutrinos and Abelian Gauge Symmetries}}, Ph.D. thesis,
  Heidelberg U., 2014.

\bibitem{Asai:2017ryy}
K.~Asai, K.~Hamaguchi and N.~Nagata, \emph{{Predictions for the neutrino
  parameters in the minimal gauged U(1)$_{L_\mu-L_\tau}$ model}},
  \href{https://doi.org/10.1140/epjc/s10052-017-5348-x}{\emph{Eur. Phys. J. C}
  {\bfseries 77} (2017) 763}
  [\href{https://arxiv.org/abs/1705.00419}{{\ttfamily 1705.00419}}].

\bibitem{Asai:2018ocx}
K.~Asai, K.~Hamaguchi, N.~Nagata, S.-Y.~Tseng and K.~Tsumura, \emph{{Minimal
  Gauged U(1)$_{L_\alpha - L_\beta}$ Models Driven into a Corner}},
  \href{https://doi.org/10.1103/PhysRevD.99.055029}{\emph{Phys. Rev. D}
  {\bfseries 99} (2019) 055029}
  [\href{https://arxiv.org/abs/1811.07571}{{\ttfamily 1811.07571}}].

\bibitem{Asai:2019ciz}
K.~Asai, \emph{{Predictions for the neutrino parameters in the minimal model
  extended by linear combination of U(1)$_{L_e-L_\mu}$, U(1)$_{L_\mu-L_\tau}$
  and U(1)$_{B-L}$ gauge symmetries}},
  \href{https://doi.org/10.1140/epjc/s10052-020-7622-6}{\emph{Eur. Phys. J. C}
  {\bfseries 80} (2020) 76} [\href{https://arxiv.org/abs/1907.04042}{{\ttfamily
  1907.04042}}].

\bibitem{Ma:2001tb}
E.~Ma and D.P.~Roy, \emph{{Anomalous neutrino interaction, muon g-2, and atomic
  parity nonconservation}},
  \href{https://doi.org/10.1103/PhysRevD.65.075021}{\emph{Phys. Rev. D}
  {\bfseries 65} (2002) 075021}
  [\href{https://arxiv.org/abs/hep-ph/0111385}{{\ttfamily hep-ph/0111385}}].

\bibitem{Baek:2001kca}
S.~Baek, N.G.~Deshpande, X.G.~He and P.~Ko, \emph{{Muon anomalous g-2 and
  gauged L(muon) - L(tau) models}},
  \href{https://doi.org/10.1103/PhysRevD.64.055006}{\emph{Phys. Rev. D}
  {\bfseries 64} (2001) 055006}
  [\href{https://arxiv.org/abs/hep-ph/0104141}{{\ttfamily hep-ph/0104141}}].

\bibitem{Foldenauer:2018zrz}
P.~Foldenauer, \emph{{Light dark matter in a gauged $U(1)_{L_\mu-L_\tau}$
  model}}, \href{https://doi.org/10.1103/PhysRevD.99.035007}{\emph{Phys. Rev.
  D} {\bfseries 99} (2019) 035007}
  [\href{https://arxiv.org/abs/1808.03647}{{\ttfamily 1808.03647}}].

\bibitem{Holst:2021lzm}
I.~Holst, D.~Hooper and G.~Krnjaic, \emph{{The Simplest and Most Predictive
  Model of Muon $g-2$ and Thermal Dark Matter}},
  \href{https://arxiv.org/abs/2107.09067}{{\ttfamily 2107.09067}}.

\bibitem{Drees:2021rsg}
M.~Drees and W.~Zhao, \emph{{$U(1)_{L_\mu-L_\tau}$ for Light Dark Matter,
  $g_\mu-2$, the $511$ keV excess and the Hubble Tension}},
  \href{https://arxiv.org/abs/2107.14528}{{\ttfamily 2107.14528}}.

\bibitem{Escudero:2019gzq}
M.~Escudero, D.~Hooper, G.~Krnjaic and M.~Pierre, \emph{{Cosmology with A Very
  Light L$_{\mu}$ \ensuremath{-} L$_{\tau}$ Gauge Boson}},
  \href{https://doi.org/10.1007/JHEP03(2019)071}{\emph{JHEP} {\bfseries 03}
  (2019) 071} [\href{https://arxiv.org/abs/1901.02010}{{\ttfamily
  1901.02010}}].

\bibitem{LDMX:2018cma}
{\scshape LDMX} collaboration, \emph{{Light Dark Matter eXperiment (LDMX)}},
  \href{https://arxiv.org/abs/1808.05219}{{\ttfamily 1808.05219}}.

\bibitem{Ezhela:2003pp}
V.V.~Ezhela, S.B.~Lugovsky and O.V.~Zenin, \emph{{Hadronic part of the muon g-2
  estimated on the sigma**2003(tot)(e+ e- ---\ensuremath{>} hadrons) evaluated
  data compilation}},  \href{https://arxiv.org/abs/hep-ph/0312114}{{\ttfamily
  hep-ph/0312114}}.

\bibitem{Zyla:2020zbs}
{\scshape Particle Data Group} collaboration, \emph{{Review of Particle
  Physics}}, \href{https://doi.org/10.1093/ptep/ptaa104}{\emph{PTEP} {\bfseries
  2020} (2020) 083C01}.

\bibitem{SHiP:2017wac}
{\scshape SHiP} collaboration, \emph{{The active muon shield in the SHiP
  experiment}},
  \href{https://doi.org/10.1088/1748-0221/12/05/P05011}{\emph{JINST} {\bfseries
  12} (2017) P05011} [\href{https://arxiv.org/abs/1703.03612}{{\ttfamily
  1703.03612}}].

\bibitem{Kim:1973he}
K.J.~Kim and Y.-S.~Tsai, \emph{{IMPROVED WEIZSACKER-WILLIAMS METHOD AND ITS
  APPLICATION TO LEPTON AND W BOSON PAIR PRODUCTION}},
  \href{https://doi.org/10.1103/PhysRevD.8.3109}{\emph{Phys. Rev. D} {\bfseries
  8} (1973) 3109}.

\bibitem{Tsai:1973py}
Y.-S.~Tsai, \emph{{Pair Production and Bremsstrahlung of Charged Leptons}},
  \href{https://doi.org/10.1103/RevModPhys.46.815}{\emph{Rev. Mod. Phys.}
  {\bfseries 46} (1974) 815}.

\bibitem{Tsai:1986tx}
Y.-S.~Tsai, \emph{{AXION BREMSSTRAHLUNG BY AN ELECTRON BEAM}},
  \href{https://doi.org/10.1103/PhysRevD.34.1326}{\emph{Phys. Rev. D}
  {\bfseries 34} (1986) 1326}.

\bibitem{Kirpichnikov:2021jev}
D.V.~Kirpichnikov, H.~Sieber, L.M.~Bueno, P.~Crivelli and M.M.~Kirsanov,
  \emph{{Probing hidden sectors with a muon beam: Total and differential cross
  sections for vector boson production in muon bremsstrahlung}},
  \href{https://doi.org/10.1103/PhysRevD.104.076012}{\emph{Phys. Rev. D}
  {\bfseries 104} (2021) 076012}
  [\href{https://arxiv.org/abs/2107.13297}{{\ttfamily 2107.13297}}].

\bibitem{Liu:2016mqv}
Y.-S.~Liu, D.~McKeen and G.A.~Miller, \emph{{Validity of the
  Weizs\"acker-Williams approximation and the analysis of beam dump
  experiments: Production of a new scalar boson}},
  \href{https://doi.org/10.1103/PhysRevD.95.036010}{\emph{Phys. Rev. D}
  {\bfseries 95} (2017) 036010}
  [\href{https://arxiv.org/abs/1609.06781}{{\ttfamily 1609.06781}}].

\bibitem{Liu:2017htz}
Y.-S.~Liu and G.A.~Miller, \emph{{Validity of the Weizs\"acker-Williams
  approximation and the analysis of beam dump experiments: Production of an
  axion, a dark photon, or a new axial-vector boson}},
  \href{https://doi.org/10.1103/PhysRevD.96.016004}{\emph{Phys. Rev. D}
  {\bfseries 96} (2017) 016004}
  [\href{https://arxiv.org/abs/1705.01633}{{\ttfamily 1705.01633}}].

\bibitem{niki:2022m}
A.~Niki, \emph{{New Gauge Bosons at ILC Beam Dump Experiment}}, {\emph{Master
  Thesis} (2022) }.

\bibitem{Bjorken:2009mm}
J.D.~Bjorken, R.~Essig, P.~Schuster and N.~Toro, \emph{{New Fixed-Target
  Experiments to Search for Dark Gauge Forces}},
  \href{https://doi.org/10.1103/PhysRevD.80.075018}{\emph{Phys. Rev. D}
  {\bfseries 80} (2009) 075018}
  [\href{https://arxiv.org/abs/0906.0580}{{\ttfamily 0906.0580}}].

\bibitem{Bauer:2018onh}
M.~Bauer, P.~Foldenauer and J.~Jaeckel, \emph{{Hunting All the Hidden
  Photons}}, \href{https://doi.org/10.1007/JHEP07(2018)094}{\emph{JHEP}
  {\bfseries 07} (2018) 094}
  [\href{https://arxiv.org/abs/1803.05466}{{\ttfamily 1803.05466}}].

\bibitem{Sakaki:2020cux}
Y.~Sakaki, Y.~Namito, T.~Sanami, H.~Iwase and H.~Hirayama,
  \emph{{Implementation of muon pair production in PHITS and verification by
  comparing with the muon shielding experiment at SLAC}},
  \href{https://doi.org/10.1016/j.nima.2020.164323}{\emph{Nucl. Instrum. Meth.
  A} {\bfseries 977} (2020) 164323}
  [\href{https://arxiv.org/abs/2004.00212}{{\ttfamily 2004.00212}}].

\bibitem{TEXONO:2009knm}
{\scshape TEXONO} collaboration, \emph{{Measurement of Nu(e)-bar -Electron
  Scattering Cross-Section with a CsI(Tl) Scintillating Crystal Array at the
  Kuo-Sheng Nuclear Power Reactor}},
  \href{https://doi.org/10.1103/PhysRevD.81.072001}{\emph{Phys. Rev. D}
  {\bfseries 81} (2010) 072001}
  [\href{https://arxiv.org/abs/0911.1597}{{\ttfamily 0911.1597}}].

\bibitem{Alekhin:2015byh}
S.~Alekhin et~al., \emph{{A facility to Search for Hidden Particles at the CERN
  SPS: the SHiP physics case}},
  \href{https://doi.org/10.1088/0034-4885/79/12/124201}{\emph{Rept. Prog.
  Phys.} {\bfseries 79} (2016) 124201}
  [\href{https://arxiv.org/abs/1504.04855}{{\ttfamily 1504.04855}}].

\bibitem{Bjorken:1988as}
J.D.~Bjorken, S.~Ecklund, W.R.~Nelson, A.~Abashian, C.~Church, B.~Lu et~al.,
  \emph{{Search for Neutral Metastable Penetrating Particles Produced in the
  SLAC Beam Dump}}, \href{https://doi.org/10.1103/PhysRevD.38.3375}{\emph{Phys.
  Rev. D} {\bfseries 38} (1988) 3375}.

\bibitem{Kamada:2015era}
A.~Kamada and H.-B.~Yu, \emph{{Coherent Propagation of PeV Neutrinos and the
  Dip in the Neutrino Spectrum at IceCube}},
  \href{https://doi.org/10.1103/PhysRevD.92.113004}{\emph{Phys. Rev. D}
  {\bfseries 92} (2015) 113004}
  [\href{https://arxiv.org/abs/1504.00711}{{\ttfamily 1504.00711}}].

\bibitem{Bellini:2011rx}
G.~Bellini et~al., \emph{{Precision measurement of the 7Be solar neutrino
  interaction rate in Borexino}},
  \href{https://doi.org/10.1103/PhysRevLett.107.141302}{\emph{Phys. Rev. Lett.}
  {\bfseries 107} (2011) 141302}
  [\href{https://arxiv.org/abs/1104.1816}{{\ttfamily 1104.1816}}].

\bibitem{Kaneta:2016uyt}
Y.~Kaneta and T.~Shimomura, \emph{{On the possibility of a search for the
  $L_\mu - L_\tau$ gauge boson at Belle-II and neutrino beam experiments}},
  \href{https://doi.org/10.1093/ptep/ptx050}{\emph{PTEP} {\bfseries 2017}
  (2017) 053B04} [\href{https://arxiv.org/abs/1701.00156}{{\ttfamily
  1701.00156}}].

\bibitem{Delahaye:2013jla}
J.-P.~Delahaye et~al., \emph{{Enabling Intensity and Energy Frontier Science
  with a Muon Accelerator Facility in the U.S.: A White Paper Submitted to the
  2013 U.S. Community Summer Study of the Division of Particles and Fields of
  the American Physical Society}},  in \emph{{Community Summer Study 2013}:
  {Snowmass on the Mississippi}}, 8, 2013
  [\href{https://arxiv.org/abs/1308.0494}{{\ttfamily 1308.0494}}].

\bibitem{Schuster:2021mlr}
P.~Schuster, N.~Toro and K.~Zhou, \emph{{Probing invisible vector meson decays
  with the NA64 and LDMX experiments}},
  \href{https://doi.org/10.1103/PhysRevD.105.035036}{\emph{Phys. Rev. D}
  {\bfseries 105} (2022) 035036}
  [\href{https://arxiv.org/abs/2112.02104}{{\ttfamily 2112.02104}}].

\end{thebibliography}\endgroup



\end{document}